\documentclass[12pt]{amsart}
\usepackage{amscd,amsmath, amsfonts, amsthm, amssymb, cite,latexsym,enumerate}
\usepackage[mathscr]{euscript}
\usepackage{epsfig}
\usepackage{fancybox}
\usepackage{verbatim}

\usepackage{color}
\usepackage[all]{xy}
\usepackage{fancyhdr}
\begin{document}



\newtheorem{theorem}{Theorem}[section]
\newcommand{\mar}[1]{{\marginpar{\textsf{#1}}}}
\numberwithin{equation}{section}
\newtheorem*{theorem*}{Theorem}
\newtheorem{prop}[theorem]{Proposition}
\newtheorem*{prop*}{Proposition}
\newtheorem{lemma}[theorem]{Lemma}
\newtheorem{corollary}[theorem]{Corollary}
\newtheorem*{conj*}{Conjecture}
\newtheorem*{corollary*}{Corollary}
\newtheorem{definition}[theorem]{Definition}
\newtheorem*{definition*}{Definition}
\newtheorem{remarks}[theorem]{Remarks}
\newtheorem*{remarks*}{remarks}
\newtheorem{rems}[theorem]{Remarks}
\newtheorem{rem}{Remark}
\newtheorem*{rem*}{Remark}
\newtheorem*{rems*}{Remarks}
\newtheorem{cor}[theorem]{Corollary}
\newtheorem{defn}[theorem]{Definition}
\newtheorem{thm}[theorem]{Theorem}

\newcommand{\isom}{V}

\newcommand{\redu}{{\mbox{\rm\tiny red}}}
\newcommand{\ph}{{\mbox{\rm\tiny ph}}}
\newcommand{\ess}{{\mbox{\rm\tiny ess}}}
\newcommand{\sym}{{\mbox{\rm\tiny sym}}}
\newcommand{\Maj}{{\mbox{\rm\tiny Maj}}}
\newcommand{\one}{{\bf 1}}
\newcommand{\nul}{{\bf 0}}
\newcommand{\inv}{{\mbox{\rm\tiny inv}}}
\newcommand{\up}{{\mbox{\rm\tiny upp}}}
\newcommand{\TR}{{\rm Tr}} 
\newcommand{\ev}{{\mbox{\rm ev}}}
\newcommand{\SF}{{\rm Sf}} 
\newcommand{\Ch}{{\rm Ch}} 
\newcommand{\Ran}{{\rm Ran}} 
\newcommand{\sgn}{{\rm sgn}} 
\newcommand{\sig}{{\rm sig}} 
\newcommand{\BM}{{\mathbb B}}
\newcommand{\CM}{{\mathbb C}}
\newcommand{\NM}{{\mathbb N}}
\newcommand{\RM}{{\mathbb R}}
\newcommand{\SM}{{\mathbb S}}
\newcommand{\TM}{{\mathbb T}}
\newcommand{\ZM}{{\mathbb Z}}
\newcommand{\PM}{{\mathbb P}}
\newcommand{\LM}{{\mathbb L}}
\newcommand{\KM}{{\mathbb K}}
\newcommand{\HM}{{\mathbb H}}
\newcommand{\DM}{{\mathbb D}}
\newcommand{\GM}{{\mathbb G}}
\newcommand{\UM}{{\mathbb U}}
\newcommand{\WM}{{\mathbb W}}
\newcommand{\FM}{{\mathbb F}}
\newcommand{\IM}{{\mathbb I}}
\newcommand{\Aa}{{\cal A}}
\newcommand{\PP}{{\bf P}}
\newcommand{\UU}{{\bf U}}
\newcommand{\pp}{{\bf p}}
\newcommand{\BB}{{\bf B}}
\newcommand{\EE}{{\bf E}}
\newcommand{\Bb}{{\cal B}}
\newcommand{\Ff}{{\cal F}}
\newcommand{\Gg}{{\cal G}}
\newcommand{\Ww}{{\cal W}}
\newcommand{\Uu}{{\cal U}}
\newcommand{\Vv}{{\cal V}}
\newcommand{\Ss}{{\cal S}}
\newcommand{\Oo}{{\cal O}}
\newcommand{\Tr}{\mbox{\rm Tr}}
\newcommand{\Tt}{{\cal T}}
\newcommand{\Rr}{{\cal R}}
\newcommand{\Nn}{{\cal N}}
\newcommand{\Mm}{{\cal M}}
\newcommand{\Cc}{{\cal C}}
\newcommand{\Jj}{{\cal J}}
\newcommand{\Ii}{{\cal I}}
\newcommand{\Ll}{{\cal L}}
\newcommand{\Kk}{{\cal K}}
\newcommand{\Hh}{{\cal H}}
\newtheorem*{not*}{Notation}
\newcommand\pa{\partial}
\newcommand\cohom{\operatorname{H}}
\newcommand\Td{\operatorname{Td}}
\newcommand\Trig{\operatorname{Trig}}
\newcommand\Hom{\operatorname{Hom}}
\newcommand\End{\operatorname{End}}
\newcommand\Ker{\operatorname{Ker}}
\newcommand\Ind{\operatorname{Ind}}
\newcommand\cker{\operatorname{coker}}
\newcommand\oH{\operatorname{H}}
\newcommand\oK{\operatorname{K}}
\newcommand\codim{\operatorname{codim}}
\newcommand\Exp{\operatorname{Exp}}
\newcommand\CAP{{\mathcal AP}}
\newcommand\T{\mathbb T}
\newcommand{\M}{\mathcal{M}}
\newcommand\ep{\epsilon}
\newcommand\te{\tilde e}
\newcommand\Dd{{\mathcal D}}

\newcommand\what{\widehat}
\newcommand\wtit{\widetilde}
\newcommand\mfS{{\mathfrak S}}
\newcommand\cA{{\mathcal A}}
\newcommand\maA{{\mathcal A}}
\newcommand\maF{{\mathcal F}}
\newcommand\maN{{\mathcal N}}
\newcommand\cM{{\mathcal M}}
\newcommand\maE{{\mathcal E}}
\newcommand\cF{{\mathcal F}}
\newcommand\maG{{\mathcal G}}
\newcommand\cG{{\mathcal G}}
\newcommand\cH{{\mathcal H}}
\newcommand\maH{{\mathcal H}}
\renewcommand\H{{\mathcal H}}
\newcommand\cO{{\mathcal O}}
\newcommand\cR{{\mathcal R}}
\newcommand\cS{{\mathcal S}}
\newcommand\cU{{\mathcal U}}
\newcommand\cV{{\mathcal V}}
\newcommand\cX{{\mathcal X}}
\newcommand\cD{{\mathcal D}}
\newcommand\cnn{{\mathcal N}}
\newcommand\wD{\widetilde{D}}
\newcommand\wL{\widetilde{L}}
\newcommand\wM{\widetilde{M}}
\newcommand\wV{\widetilde{V}}
\newcommand\Ee{{\mathcal E}}
\newcommand{\npartial}{\slash\!\!\!\partial}
\newcommand{\Heis}{\operatorname{Heis}}
\newcommand{\Solv}{\operatorname{Solv}}
\newcommand{\Spin}{\operatorname{Spin}}
\newcommand{\SO}{\operatorname{SO}}
\newcommand{\ind}{\operatorname{ind}}
\newcommand{\Index}{\operatorname{Index}}
\newcommand{\ch}{\operatorname{ch}}
\newcommand{\rank}{\operatorname{rank}}
\newcommand{\G}{\Gamma}
\newcommand{\HK}{\operatorname{HK}}
\newcommand{\Dix}{\operatorname{Dix}}

\newcommand{\tM}{\tilde{M}}  
\newcommand{\tS}{\tilde{S}}
\newcommand{\tH}{\tilde{\mathcal H}}
\newcommand{\tg}{\tilde{g}}
\newcommand{\tx}{\tilde{x}}
\newcommand{\ty}{\tilde{y}}
\newcommand{\ox}{\otimes}


\newcommand{\abs}[1]{\lvert#1\rvert}
 \newcommand{\A}{{\mathcal A}}
        \newcommand{\D}{{\mathcal D}}
\newcommand{\HH}{{\mathcal H}}
        \newcommand{\LL}{{\mathcal L}}
        \newcommand{\B}{{\mathcal B}}
        \newcommand{\K}{{\mathcal K}}
\newcommand{\oo}{{\mathcal O}}
         \newcommand{\Pp}{{\mathcal P}}
         \newcommand{\Qq}{{\mathcal Q}}
        \newcommand{\s}{\sigma}
        \newcommand{\coker}{{\mbox coker}}
        \newcommand{\dd}{|\D|}
        \newcommand{\n}{\parallel}
\newcommand{\bma}{\left(\begin{array}{cc}}
\newcommand{\ema}{\end{array}\right)}
\newcommand{\bba}{\left(\begin{array}{ccc}}
\newcommand{\eba}{\end{array}\right)}
\newcommand{\bvba}{\left(\begin{array}{cccc}}
\newcommand{\evba}{\end{array}\right)}

\newcommand{\bca}{\left(\begin{array}{c}}
\newcommand{\eca}{\end{array}\right)}
\newcommand{\sr}{\stackrel}
\newcommand{\da}{\downarrow}
\newcommand{\tD}{\tilde{\D}}
        \newcommand{\R}{\mathbb R}
        \newcommand{\C}{\mathbb C}
        \newcommand{\h}{\mathbb H}
\newcommand{\Z}{\mathcal Z}
\newcommand{\N}{\mathbb N}
\newcommand{\tto}{\longrightarrow}
\newcommand{\ZZ}{{\mathcal Z}}
\newcommand{\ben}{\begin{displaymath}}
        \newcommand{\een}{\end{displaymath}}
\newcommand{\be}{\begin{equation}}
\newcommand{\ee}{\end{equation}}
        \newcommand{\bean}{\begin{eqnarray*}}
        \newcommand{\eean}{\end{eqnarray*}}
\newcommand{\nno}{\nonumber\\}
\newcommand{\bea}{\begin{eqnarray}}
        \newcommand{\eea}{\end{eqnarray}}
\newcommand{\x}{\times}

\newcommand{\Ga}{\Gamma}
\newcommand{\e}{\epsilon}
\renewcommand{\L}{\mathcal{L}}
\newcommand{\supp}[1]{\operatorname{#1}}
\newcommand{\norm}[1]{\parallel\, #1\, \parallel}
\newcommand{\ip}[2]{\langle #1,#2\rangle}
\newcommand{\nc}{\newcommand}
\nc{\gf}[2]{\genfrac{}{}{0pt}{}{#1}{#2}}
\nc{\mb}[1]{{\mbox{$ #1 $}}}
\nc{\real}{{\mathbb R}}
\nc{\comp}{{\mathbb C}}
\nc{\ints}{{\mathbb Z}}
\nc{\Ltoo}{\mb{L^2({\mathbf H})}}
\nc{\rtoo}{\mb{{\mathbf R}^2}}
\nc{\slr}{{\mathbf {SL}}(2,\real)}
\nc{\slz}{{\mathbf {SL}}(2,\ints)}
\nc{\su}{{\mathbf {SU}}(1,1)}
\nc{\so}{{\mathbf {SO}}}
\nc{\hyp}{{\mathbb H}}
\nc{\disc}{{\mathbf D}}
\nc{\torus}{{\mathbb T}}
\newcommand{\tk}{\widetilde{K}}
\newcommand{\boe}{{\bf e}}\newcommand{\bt}{{\bf t}}
\newcommand{\vth}{\vartheta}
\newcommand{\CGh}{\widetilde{\CG}}
\newcommand{\db}{\overline{\partial}}
\newcommand{\tE}{\widetilde{E}}
\newcommand{\tr}{{\rm tr}}
\newcommand{\ta}{\widetilde{\alpha}}
\newcommand{\tb}{\widetilde{\beta}}
\newcommand{\txi}{\widetilde{\xi}}
\newcommand{\hV}{\hat{V}}
\newcommand{\IC}{\mathbf{C}}
\newcommand{\IZ}{\mathbf{Z}}
\newcommand{\IP}{\mathbf{P}}
\newcommand{\IR}{\mathbf{R}}
\newcommand{\IH}{\mathbf{H}}
\newcommand{\IG}{\mathbf{G}}
\newcommand{\IS}{\mathbf{S}}
\newcommand{\CC}{{\mathcal C}}
\newcommand{\CS}{{\mathcal S}}
\newcommand{\CG}{{\mathcal G}}
\newcommand{\CL}{{\mathcal L}}
\newcommand{\CO}{{\mathcal O}}
\nc{\ca}{{\mathcal A}}
\nc{\cag}{{{\mathcal A}^\Gamma}}
\nc{\cg}{{\mathcal G}}
\nc{\chh}{{\mathcal H}}
\nc{\ck}{{\mathcal B}}
\nc{\cd}{{\mathcal D}}
\nc{\cl}{{\mathcal L}}
\nc{\cm}{{\mathcal M}}
\nc{\cn}{{\mathcal N}}
\nc{\cs}{{\mathcal S}}
\nc{\cz}{{\mathcal Z}}
\nc{\sind}{\sigma{\rm -ind}}

\newcommand\clFN{{\mathcal F_\tau(\mathcal N)}}       
\newcommand\clKN{{\mathcal K_\tau(\mathcal N)}}       
\newcommand\clQN{{\mathcal Q_\tau(\mathcal N)}}       %
\newcommand\tF{\tilde F}
\newcommand\clA{\mathcal A}
\newcommand\clH{\mathcal H}
\newcommand\clN{\mathcal N}
\newcommand\Del{\Delta}
\newcommand\g{\gamma}
\newcommand\eps{\varepsilon}
\newcommand\vf{\varphi}
\newcommand\E{\mathcal E}

\newcommand{\CDA}{\mathcal{C_D(A)}} 
\newcommand{\dslash}{{\pa\mkern-10mu/\,}}

\newcommand{\sepword}[1]{\quad\mbox{#1}\quad} 

\nc{\nt}{\newtheorem}
\nc{\bra}{\langle}
\nc{\ket}{\rangle}
\nc{\cal}{\mathcal}
\nc{\frk}{\mathfrak}

\parindent=0.0in

 \title{Spectral flow for skew-adjoint Fredholm operators}
\footnote{\uppercase{T}his work is partially supported by the Australian Research
Council, the NSERC, and the DFG.}

\author{Alan L. CAREY}
\address{Mathematical Sciences Institute,
 Australian National University\\
 Canberra ACT, 0200 Australia, and
 \newline
 School of Mathematics and Applied Statistics\\
University of Wollongong\\
Wollongong NSW, 2500 Australia
\newline
 e-mail: alan.carey@anu.edu.au\\}

\author{John PHILLIPS}
\address{Department of Mathematics and Statistics\\
University of Victoria\\
Victoria BC, Canada
\newline
e-mail: johnphil@uvic.ca}

\author{Hermann SCHULZ-BALDES}
\address{
Department Mathematik, Friedrich-Alexander-Universit\"at Erlangen-N\"urnberg, 
Germany
\newline
e-mail: schuba@mi.uni-erlangen.de}

\begin{abstract} 
An analytic definition of a $\ZM_2$-valued spectral flow for paths of real skew-adjoint Fredholm operators is given. It counts the parity of the number of changes in the orientation of the eigenfunctions at eigenvalue crossings through $0$ along the path. The $\ZM_2$-valued spectral flow is shown to satisfy a concatenation property and homotopy invariance, and it provides an isomorphism on the fundamental group of the real skew-adjoint Fredholm operators. Moreover, it is connected to a $\ZM_2$-index pairing for suitable paths. Applications concern the zero energy bound states at defects in a Majorana chain and a spectral flow interpretation for the $\ZM_2$-polarization in these models.
\end{abstract}

\maketitle


\section{Introduction}

The main objective of this paper is to construct a $\ZM_2$-valued spectral flow for paths of skew-adjoint Fredholms on a real Hilbert space. Our justification for using the term
`spectral flow' for the spectral invariant defined here is that it satisfies the three properties that can be used to axiomatize \cite{Les} the spectral flow for the self-adjoint Fredholm operators on a complex Hilbert space \cite{APS,Ph}, namely:

\vspace{.1cm}

(i) normalisation, \hspace{1cm} (ii) concatenation,\hspace{1cm} (iii) homotopy invariance.

\vspace{.1cm}

In Section~\ref{sec-FiniteDim} this will first be achieved for spectral flow along straight line paths in finite dimensions. The correct definition simply counts the number of orientation changes of the eigenfunctions at eigenvalue crossings at $0$ modulo $2$. Hence the $\ZM_2$-spectral flow is {\em not} of purely spectral nature, but also depends on the eigenfunctions, as explained on a particularly simple example in Section~\ref{sec-FiniteDim}. For the extension to Fredholm operators we then follow in Section~\ref{sec-def} the analytic approach to the complex spectral flow for paths of self-adjoint Fredholm operators on a complex Hilbert space as described in \cite{Ph},  namely a partitioning argument is used allowing to restrict to the finite dimensional case. In the complex case, this circumvents considerable technical difficulties linked to the topologists' intersection number approach ({\it e.g.} \cite{G} based on \cite{Ruget}) and leads to computable formulas \cite{CP}. In the present context it allows to show relatively directly that the $\ZM_2$-valued spectral flow can be calculated, similarly to the complex spectral flow \cite{BCPRSW}, as a sum of index type contributions (Section~\ref{sec-AlternativeForm}), provided the appropriate notion of index is used. This turns out to be the \lq\lq mod 2" index map on a subgroup of the orthogonal group which was introduced in \cite{CO} and is reviewed in Section~\ref{sec-prelimOrtho}. Finally an index formula is proved in Section~\ref{sec-index} which connects the $\ZM_2$-valued spectral flow of certain paths in the skew-adjoint operators on a real Hilbert space to the $\ZM_2$-index of an associated Toeplitz operator on the complexification. All of this is illustrated in Section~\ref{sec-ex} by an explicit example given by a matrix-valued shift operator which can be considered to be the analogue in real Hilbert space of the standard Toeplitz operator in the complex case. This example can be viewed as the canonical non-trivial example of $\ZM_2$-valued spectral flow. 

\vspace{.2cm}

Next, let us place the $\ZM_2$-valued spectral flow into the perspective of the work of Atiyah and Singer \cite{AS} on the classifying spaces for real $K$-theory. One of these spaces is the set $\Ff^1(\Hh_\RM)$ of skew-adjoint Fredholm operators on a separable real Hilbert space $\Hh_\RM$. Its homotopy groups are known to be $8$-periodic and given by
\begin{equation}
\label{eq-homotopygroups}
\begin{tabular}{|c||c|c|c|c|c|c|c|c|}
\hline
$i$ & $0$ & $1$ & $2$ & $3$ & $4$ & $5$ & $6$ & $7$  \\
\hline
$\pi_{i}(\Ff^1(\Hh_\RM))$  &$\ZM_2$ & $\ZM_2$ & $0$ & $2\,\ZM$ & $0$ & $0$ & $0$ & $\ZM$ 
\\
\hline
\end{tabular}
\end{equation}
The two components of $\Ff^1(\Hh_\RM)$, that is $\pi_{0}(\Ff^1(\Hh_\RM))$, can be read off the parity of the kernel dimension (see  \cite{AS} and Section~\ref{sec-prelim} below). One of the contribution of this paper is to show that $\pi_{1}(\Ff^1(\Hh_\RM))\cong\ZM_2$ is detected by the $\ZM_2$-valued spectral flow, which actually provides an explicit isomorphism (see Section~\ref{sec-index}). We do not attempt here, however, to  obtain topological formulas for the $\ZM_2$-valued spectral flow using real analogues of the Atiyah-Singer index theorem.  To our knowledge this is an unresolved problem. Furthermore, we do not study $\pi_{1}(\Ff^7(\Hh_\RM))\cong\ZM_2$ as a spectral flow here. Let us also note that several of the results below can be rewritten using Clifford valued indices as described in Section~III.10 of \cite{LM}, but this will not be spelled out in any detail.

\vspace{.2cm}

While this paper focuses on the mathematical questions addressed above,  our motivation comes from  the use of real $K$-theory in mathematical physics.  We present two examples of applications, both to the  theory of topological insulators. The interested reader should see Section~\ref{sec-TopIns} for details. In this context the importance of a real spectral flow has recently been highlighted \cite{Witten}. Let us point out that another notion of $\ZM_2$-spectral flow more closely linked to the complex spectral flow under a particular symmetry condition was investigated in \cite{DS,DS2}. This is not connected to the one studied in the present paper.

\section{$\ZM_2$-valued spectral flow for linear paths in finite dimension}
\label{sec-FiniteDim}

This section defines the $\ZM_2$-valued spectral flow associated to a linear (or straight line) path $t\in[0,1]\mapsto T_t=(1-t)T_0+t\,T_1$ of skew-adjoint operators on a finite dimensional Hilbert space $\Hh_\RM$. As a motivation, let us begin with $\Hh_\RM=\RM^2$ and consider two paths, one linear and a second non-analytic path $t\in[0,1]\mapsto \tilde T_t$ of skew-adjoint matrices:
\begin{equation}
\label{eq-2x2example}
T_t\;=\;(2t-1)
\begin{pmatrix}
0 & -1 \\ 1 & 0
\end{pmatrix}
\;,
\qquad
\tilde T_t\;=\;|2t-1|
\begin{pmatrix}
0 & -1 \\ 1 & 0
\end{pmatrix}
\;.
\end{equation}
The spectra of $T_t$ and $\tilde T_t$ as complex operators are $\sigma(T_t)=\sigma(\tilde T_t)=\{(1-2t)\imath,(2t-1)\imath\}$ with $\imath=\sqrt{-1}$ so that both eigenvalues form a crossing with a double degenerate kernel at $t=\frac{1}{2}$, and the associated complex spectral flow (in any possible sense, {\it e.g.} of \cite{APS,Ph}) vanishes. Nevertheless, there is a difference between the two paths. In fact, for $\tilde T_t$, one can consider the homotopy $s\in[0,1]\mapsto \tilde T_t(s)$ of paths of skew-adjoints given by
$$
\tilde T_t(s)\;=\;|2ts-1|
\begin{pmatrix}
0 & -1 \\ 1 & 0
\end{pmatrix}
\;.
$$
Then $\tilde T_t(1)=\tilde T_t$, while $\tilde T_t(0)$ is a constant path with spectrum $\sigma(\tilde T_t(0))=\{-\imath,\imath\}$ which is actually the straight-line path between $\tilde T_0$ and $\tilde T_1$. Consequently the spectral crossing of  the path $\tilde T_t$ can be homotopically lifted. On the other hand, it is impossible to lift the kernel of $T_t$. This defect is encoded in the eigenfunctions as follows. Viewing $T_0$ and $T_1$ as non-degenerate skew-symmetric bilinear forms, results from linear algebra imply that there exists a real invertible matrix $A$ such that 
$$
T_1\;=\;A^*T_0A
\;.
$$
Actually, here $A=\binom{0\;1}{1\;0}$ which exchanges the eigenvectors of the upper and lower branch of $T_t$ at $t=\frac{1}{2}$. This is reflected by the sign of $\det(A)$ and this sign is by definition the $\ZM_2$-valued spectral flow $\SF_2(T_0,T_1)$ between the points $T_0$ and $T_1$ along the straight line path. This actually is true also for the complex spectral flow \cite{BCPRSW} (which is simply equal to the difference of the signatures of the end points).  Let us stress again that due to the above, this $\ZM_2$-valued spectral flow is {\em not} only determined by the spectrum of the path, but rather depends on the eigenfunctions as well. However, we will show further below that a path having vanishing kernel throughout necessarily has a trivial $\ZM_2$-valued spectral flow.  Another important difference between the two cases in \eqref{eq-2x2example} is that $T_t$ is analytic in $t$, while $\tilde T_t$ is not (this will {also} be elaborated upon further down). After these words of motivation, we can go on to the formal definition of the $\ZM_2$-spectral flow. 

\begin{definition}
\label{def-SF2finite}
Suppose given two skew-adjoint operators $T_0$ and $T_1$ on a finite dimensional real Hilbert space $\Hh_\RM$ with {the same} minimal kernel dimension, namely their kernel has dimension equal to $\dim(\Hh_\RM)\,\mbox{\rm mod}\;2$. Given an invertible $A$ such that $T_1=A^*T_0A$, the $\ZM_2$-valued spectral flow (along the straight line path) from {an invertible} $T_0$ to {an invertible} $T_1$ is defined by
$$
\SF_2(T_0,T_1)
\;=\;
\sgn(\det(A))\;\in\;\ZM_2
\;.
$$
{If $T_0$  has a one-dimensional kernel one first rotates it onto the kernel of $T_1$ and then applies the above definition on the orthogonal complement.}
\end{definition}

In the above definition, $\ZM_2$ appears as the mod $2$ dimension of a vector space and also as the sign of a determinant, hence both as the additive group $\{0,1\}$ and as the multiplicative group $\{1,-1\}$. While we freely identify these sets (namely $0$ with $1$, and $1$ with $-1$), both group structures $(\ZM_2,+)$ and $(\ZM_2,\cdot)$ will play a role in the following. As this is at the heart of the matter of the paper, we will be more careful about this point than may seem neessary to some readers.
 
\begin{lemma}
\label{lemma-SF2welldef}
The $\ZM_2$-valued spectral flow $\SF_2(T_0,T_1)$ is well-defined.
\end{lemma}
 
\noindent {\bf Proof.} It has to be checked that the definition is independent of the choice of $A$. Indeed, for any orthogonal $O$ commuting with $T_0$ one also has $T_1=(OA)^*T_0(OA)$. If $\dim(\Hh_\RM)$ is even and $T_0$ is a complex structure, then $O$ lies in the associated symplectic group and therefore has determinant $1$ so that $\sgn(\det(OA))=\sgn(\det(A))$. If $T_0$ is not a complex structure, one can homotopically deform it to one using spectral calculus. Along this path, the sign of the determinant of $A$ does not change. In the case of odd $\dim(\Hh_\RM)$ {and after having rotated the kernel of $T_0$ onto that of $T_1$}, the above argument in even dimension applies.
\hfill $\Box$

\vspace{.2cm}

The definition also directly implies 

\begin{lemma}
\label{lemma-SF2basic}
For any invertible matrices $B,C$ with $\det(C)>0$, one has
$$
\SF_2(T_0,T_1)
\;=\;
\SF_2(T_1,T_0)
\;=\;
\SF_2(BT_0B^*,BT_1B^*)
\;=\;
\SF_2(CT_0C^*,T_1)
\;.
$$
Moreover, if $T'_0$ and $T'_1$ are skew-adjoints on another finite-dimensional real Hilbert space $\Hh'_\RM$,
$$
\SF_2(T_0\oplus T'_0,T_1\oplus T'_1)
\;=\;
\SF_2(T_0,T_1)
\,+\,
\SF_2(T'_0,T'_1)
\;,
$$
with addition modulo $2$ in $(\ZM_2,+)$.
\end{lemma}
 

The following proposition indicates one use of the $\ZM_2$-valued spectral flow.

\begin{prop}
\label{prop-Z2obstruction}
Let $T_0$ and $T_1$ be skew-adjoint operators on a finite dimensional Hilbert space with minimal kernel dimension. Let there exist a path $t\in[0,1]\mapsto T_t$  from $T_0$ to $T_1$ with constant minimal kernel dimension of $T_t$ for all $t$. Then $\SF_2(T_0,T_1)=0$.
\end{prop}

\noindent {\bf Proof.} We consider only the case of an even dimensional real Hilbert space. The path provides  a continuous path $J_t=T_t|T_t|^{-1}$ of complex structures, and there thus exists a continuous path of $t\in[0,1]\mapsto O_t$ such that $J_t=O_t^*J_0O_t$. As $A=O_1$ is path connected to the identity, it follows that $\det(O_t)=1$ and thus $\SF_2(T_0,T_1)=0$. 
\hfill $\Box$

\vspace{.2cm}


Next let us discuss the so-called concatenation property of the $\ZM_2$-valued spectral flow. 

\begin{prop}
\label{prop-Z2concat}
For skew-adjoint operators $T_0$, $T_1$, $T_2$ having minimal kernel dimensions,
\begin{equation}
\label{eq-concat}
\SF_2(T_0,T_2)
\;=\;
\SF_2(T_0,T_1)
\;+\;
\SF_2(T_1,T_2)
\;,
\end{equation}
with addition modulo $2$ in $(\ZM_2,+)$.
\end{prop}

\noindent {\bf Proof.}  
If $T_1=A^*T_0A$  and $T_2=B^*T_1B$ for invertibles $A,B$, then $T_2=(AB)^*T_0(AB)$. Taking determinants the claim  follows.
\hfill $\Box$

\vspace{.2cm}

Let us note that we always took care to suppose that the end points $T_0$ and $T_1$ of the path have minimal kernel dimension. Indeed, this fixes the invertible $A$ in  $T_1=A^*T_0A$ on the kernel. If the kernels of $T_0$ and $T_1$ have different dimension, no such invertible $A$ exists. If they are of same dimension, then such an $A$ exists. However, the sign of $\det(A)$ depends on the choice of $A$ (provided that the kernel dimension is larger than $1$). One way out may seem to modify $T_0$ and $T_1$ by adding skew-adjoint perturbations $W_0$ and $W_1$ on the kernels such that the kernel dimension of $T_0+W_0$ and $T_1+W_1$ is minimal so that the above definition applies. Again, one readily checks that the outcome depends on the choices of $W_0$ and $W_1$. In conclusion, there is no reasonable definition of the $\ZM_2$-valued spectral flow if the kernel dimension of the end points is not minimal.

\vspace{.2cm}

On the other hand, this issue is not of importance for the concatenation of a subdivision of a path $t\in[0,2]\mapsto T_t$ with $T_0$ and $T_2$ having minimal kernel dimension, namely if $T_1$ does not have minimal kernel dimension. Then one can add a skew-adjoint perturbation $W_1$ on the kernel of $T_1$ such that $T_1+W_1$ has minimal kernel dimension. Now  \eqref{eq-concat} holds if $T_1$ is replaced by $T_1+W_1$. This is independent of the choice of $W_1$ because the two modifications cancel out. This fact is of considerable importance for the definition of the $\ZM_2$-valued spectral flow for arbitrary paths in infinite dimension in Section~\ref{sec-def} below. 

\vspace{.2cm}

The final preparations in finite dimension concerns the definition of a $\ZM_2$-valued spectral flow for operators on different real Hilbert spaces. This is needed to identify suitable spectral subspaces in Section~\ref{sec-def} below.

\begin{prop}
\label{prop-Z2ChangeSpace}
Let $\Ee$, $\Ee'$ and $\Ee''$ be three real Hilbert spaces of the same finite dimension and let $T,T',T''$ (resp.) be skew-adjoint operators with minimal kernel dimension on these spaces. Further let $\isom:\Ee\to\Ee'$, $\isom':\Ee'\to\Ee''$ and $\isom'':\Ee''\to\Ee$ be three isomorphisms. If $\|\isom^*\isom-{\one_\Ee}\|<1$,
$$
\SF_2(T,\isom^*T'\isom)
\;=\;
\SF_2(\isom T\isom^*,T')
\;.
$$
If $\|\isom''((\isom')^*)^{-1}\isom-{\one_\Ee}\|<1$, then, with addition modulo $2$ in $(\ZM_2,+)$,
$$
\SF_2(T,\isom''T''(\isom'')^*)
\;=\;
\SF_2(T,\isom^*T'\isom)
\;+\;
\SF_2(T',(\isom')^*T''\isom')
\;.
$$
\end{prop}

\noindent {\bf Proof.}  By Lemma~\ref{lemma-SF2basic}, $\SF_2(\isom T\isom^*,T')=\SF_2(\isom^*\isom T(\isom^*\isom)^*,\isom^*T'\isom)$. But $\|{\one_\Ee}-\isom^*\isom\|<1$ implies that $s\in[0,1]\mapsto {\one_\Ee}-s({\one_\Ee}-\isom^*\isom)$ is a path of invertibles connecting $\isom^*\isom$ to ${\one_\Ee}$. Now the last equality of Lemma~\ref{lemma-SF2basic} implies the first claim. Next let $A,A'$ be invertibles such that $A^*TA= \isom^*T'\isom$ and $(A')^*T'A'= (\isom')^*T''\isom'$. Then $BTB^*= \isom''T''(\isom'')^*$ for 
$$
B
\;=\;
\isom''((\isom')^*)^{-1}(A')^*(\isom^*)^{-1}A^*
\;=\;
[\isom''((\isom')^*)^{-1}\isom][\isom^{-1}(A')^*(\isom^{-1})^*][A^*]
\;.
$$
Now by the same argument as above the factor in the first bracket has positive determinant by assumption. The other two factors  have the same signs as $\det(A')$ and $\det(A)$.
\hfill $\Box$

\vspace{.2cm}

Finally, we prove criteria which assure the hypothesis in Proposition~\ref{prop-Z2ChangeSpace}.

\begin{prop}
\label{prop-Z2ChangeSpaceCriteria}
Let $\Ee$, $\Ee'$ be subspaces of a real Hilbert space $\Hh_\RM$, possibly of infinite dimension. Let $Q,Q'$ be the orthogonal projections on $\Ee$, $\Ee'$ respectively, and let $\isom:\Ee\to\Ee'$ be defined by $\isom v=Q'v$. Suppose that for some $\epsilon<{\frac{1}{4}}$
$$
\|Q-Q'\|\,<\,\epsilon\;.
$$
Then $\isom$ is an isomorphism with $\|\isom\|\leq 1$ and $\|\isom^{-1}\|< 1+2\epsilon$. One has
$$
\|\isom^*\isom-{\one_{\Ee}}\|\,<\,2\epsilon
\;,
\qquad
\|\isom\isom^*-{\one_{\Ee'}}\|\,<\,2\epsilon
\;.
$$
Let now $\Ee''$ be a third subspace with orthogonal projection $Q''$, and let $\isom':\Ee'\to\Ee''$ and $\isom'':\Ee''\to\Ee$ be defined by $\isom'v'=Q''v'$ and $\isom''v''=Qv''$. Suppose that, moreover, 
$$
\|Q'-Q''\|\,<\,\epsilon\;,
\qquad
\|Q''-Q\|\,<\,\epsilon
\;.
$$
Then $\isom,\isom',\isom''$ are isomorphisms and 
$$
\|\isom''((\isom')^*)^{-1}\isom-{\one_{\Ee}}\|\;<\;{6}\epsilon
\;.
$$
\end{prop}

\noindent {\bf Proof.} {Let us extend $V$ to a linear operator $W$ on $\Hh_\RM$ by setting $Ww=0$ for $w\in\Ee^\perp$. Similar $W'$ and $W''$ are extensions of $V'$ and $V''$.  Then $\|(W-Q)v\|=\|(W-Q)Qv\|\leq \epsilon \|Qv\|\leq \epsilon\|v\|$ so that $\|W-Q\|\leq\epsilon$ and thus also $\|W^*-Q\|\leq\epsilon$. As $W^*W-Q=W^*(W-Q)+(W^*-Q)Q$, one thus has $\|W^*W-Q\|\leq 2\epsilon$. Restriction shows $\|\isom^*\isom-\one_\Ee\|\leq 2\epsilon$ and similarly $\|\isom\isom^*-\one_{\Ee'}\|\leq 2\epsilon$.}  In particular, $\isom$ is bijective {from $\Ee$ to $\Ee'$}. Next 
$$
\|\isom v\|^2\;=\;\|v\|^2\,-\,v^*({\one_\Ee}-\isom^*\isom)v\;>\;(1-2\epsilon)\|v\|^2
\;.
$$
Choosing $v=\isom^{-1}w$ this implies $\|\isom^{-1}\|<(1-2\epsilon)^{-\frac{1}{2}}\leq 1+2\epsilon$ {for $\epsilon\leq \frac{1}{4}$}. Finally
{
$$
\|V'-(V'^*)^{-1}\|\;=\;\|(V'V'^*-\one_{\Ee''})(V'^*)^{-1}\|\;\leq\;2\epsilon(1+2\epsilon)\;\leq\;3\epsilon
\;,
$$
so that
}
\begin{align*}
\|\isom''((\isom')^*)^{-1}\isom-{\one_\Ee}\|\;
&
{ \leq\;
3\epsilon\;+\;
\|\isom''\isom'\isom-\one_\Ee\|
}
\\
& 
{  =\;
3\epsilon\;+\;
\|W''W'W-Q\|}
\\
&
{
 =\;
3\epsilon\;+\;
\|W''W'W-Q''W'W+W'W-Q'W+W-Q\|
\;\leq\;
6\epsilon
\;,
}
\end{align*}
which implies the last claim.
\hfill $\Box$

\section{Preliminaries on skew-adjoint Fredholm operators}
\label{sec-prelim}

Let $\Bb(\Hh_\RM)$ and $\Kk(\Hh_\RM)$ be the bounded and compact $\RM$-linear operators on a separable real Hilbert space $\Hh_\RM$. The $\CM$-linear operators on its complexification $\Hh_\CM=\CM\otimes\Hh_\RM$ are denoted by $\Bb(\Hh_\CM)$ and $\Kk(\Hh_\CM)$. The canonical complex conjugation $\Cc$ on $\Hh_\CM$ is given by $\Cc(\lambda\psi)=\overline{\lambda}\psi$ where $\lambda\in\CM$ and $\psi\in\Hh_\RM$. For $T\in\Bb(\Hh_\CM)$ we also introduce the notations $\overline{T}=\Cc T \Cc$ and $T^t=(\overline{T})^*$ for the complex conjugate and the transpose. Note that both of these operators are $\CM$-linear, even though $\Cc$ is anti-linear. An operator $T\in\Bb(\Hh_\CM)$ is called real if $\overline{T}=T$. The spectrum $\sigma(T)$ of every real operator $T\in\Bb(\Hh_\CM)$  satisfies $\overline{\sigma(T)}=\sigma(T)$. An operator $T\in\Bb(\Hh_\CM)$ is called skew-adjoint if $T^*=-T$. The spectrum of skew-adjoint operators lies on the imaginary axis, that is $\sigma(T)\subset\imath\,\RM$.

\vspace{.2cm}

Given $T\in\Bb(\Hh_\RM)$, an associated $\CM$-linear operator also denoted by $T$ is defined by $T(\lambda\psi)=\lambda T\psi$. This operator is real. Conversely, every real operator on $\Hh_\CM$ can be restricted to $\Hh_\RM$ and this restriction is clearly $\RM$-linear. Thus
$$
\Bb(\Hh_\RM)\;\cong\;
\left\{T\in\Bb(\Hh_\CM)\;|\;\overline{T}=T\right\}
\;,
\qquad
\Kk(\Hh_\RM)\;\cong\;
\left\{K\in\Kk(\Hh_\CM)\;|\;\overline{K}=K\right\}
\;.
$$
The spectrum $\sigma(T)$ of $T\in\Bb(\Hh_\RM)$ is always understood to be the spectrum of the complexification of $T$. In particular, the spectrum of every operator $T\in\Bb(\Hh_\RM)$ is invariant under complex conjugation and for a real skew-adjoint $T$ this implies $\sigma(T)=-\sigma(T)$.

\vspace{.2cm}

For $\KM=\RM$ or $\KM=\CM$, let next $\Qq(\Hh_\KM)=\Bb(\Hh_\KM)/\Kk(\Hh_\KM)$ be the Calkin algebra and $\pi:\Bb(\Hh_\KM)\to\Qq(\Hh_\KM)$ the canonical projection. Then the Fredholm operators $\Ff(\Hh_\KM)$ are those operators $T\in\Bb(\Hh_\KM)$ with invertible $\pi(T)\in\Qq(\Hh_\KM)$. Here the main object of study are the real skew-adjoint Fredholm operators
$$
\Ff^1(\Hh_\RM)
\;=\;
\left\{T\in\Ff(\Hh_\RM)\;|\;T^*=-T\right\}
\;\cong\;
\left\{T\in{\Ff}(\Hh_\CM)\;|\;\overline{T}=T\mbox{ and }T^*=-T\right\}
\;.
$$
The notation $\Ff^1(\Hh_\RM)$ is taken from the seminal paper \cite{AS} (although in their notation there is  a supplementary ${}\hat{}\ $). In \cite{AS} it is shown that $\Ff^1(\Hh_\RM)$ is one of the classifying spaces for real $K$-theory. As for any skew-adjoint operator $T\in\Bb(\Hh_\RM)$, the operator $\imath T\in\Bb(\Hh_\CM)$ is self-adjoint,  and spectral calculus is readily available. The essential spectrum is $\sigma_\ess(T)=\sigma(\pi(T))$.  One has $T\in\Ff^1(\Hh_\RM)$ if and only if $0\not\in\sigma_\ess(T)$. Furthermore,  $\Ff^1(\Hh_\RM)$ has two connected components which are distinguished by
\begin{equation}
\label{eq-Z2skew}
\Ind_1(T)
\;=\;
\dim_\RM(\Ker_\RM(T))\;\mbox{mod }2
\;\in\;\ZM_2
\;.
\end{equation}
Again, the subindex $1$ on $\Ind_1$ is in agreement with the notations of \cite{AS}. If $T$ is viewed as a real operator on the complexified Hilbert space, it is also given by
$$
\Ind_1(T)
\;=\;
\dim_\CM(\Ker_\CM(T))\;\mbox{mod }2
\;.
$$
Let us briefly recall why this is a well-defined homotopy invariant. Indeed, using spectral calculus one can contract all positive and negative imaginary spectrum to one point $\imath$ and $-\imath$, and then successively lift the degeneracy of the kernel by rank $2$ perturbations, until the dimension of the kernel is either $0$ or $1$.  Furthermore, Atiyah and Singer showed that $\Ff^1(\Hh_\RM)$ has the homotopy type as the (inductive limit) group $O$ of orthogonal matrices, namely the homotopy groups are given by $\pi_n(\Ff^1(\Hh_\RM))=\pi_{n}(O)$, see \eqref{eq-homotopygroups}. The standard example of an operator in the component with odd dimensional kernel is 
$$
T\;=
\begin{pmatrix}
0 &- S
\\
S^* & 0
\end{pmatrix}
\;,
\qquad
\mbox{\rm on }\ell^2_\RM(\NM)\otimes\RM^2
\;,
$$
where $S$ is the unilateral right shift on $\ell^2_\RM(\NM)$ with one-dimensional cokernel. 



\section{Definition and basic properties of the $\ZM_2$-valued spectral flow}
\label{sec-def}

\begin{figure}
\centering
\hspace{-3.2cm}
  \begin{minipage}[b]{0.4\textwidth}
\includegraphics[height=5.65cm]{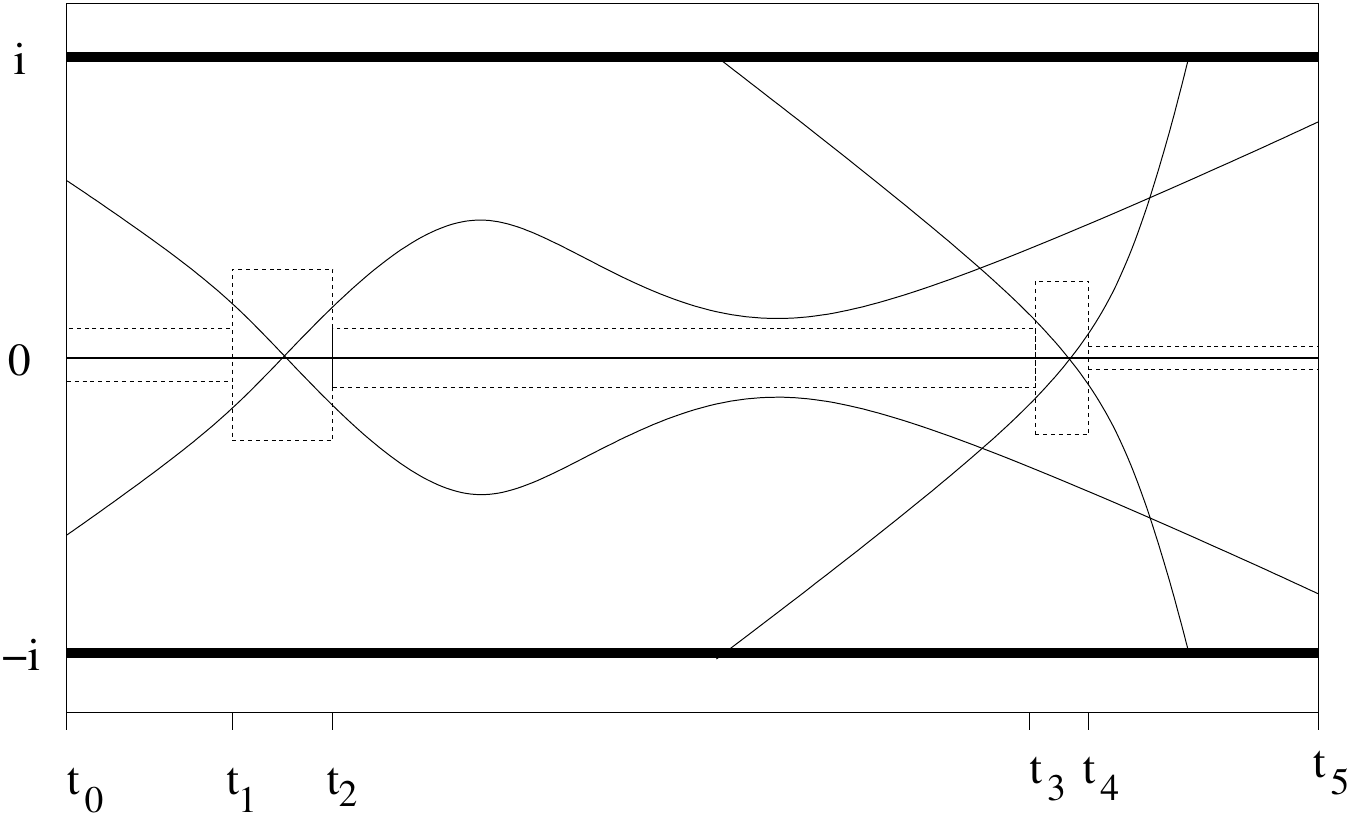}
  \end{minipage}
\caption{\it 
Schematic representation of the partition used for the definition of the $\ZM_2$-valued spectral flow. The vertical axis is the imaginary spectral axis of the operators $T_t$ which have essential spectrum $\{-\imath,\imath\}$. The values $a_n$ can be read off the heights of the dotted boxes.
\label{fig-schematic}
}
\end{figure}

Let $t\in [0,1]\mapsto T_t\in \Ff^1(\Hh_\RM)$ be a continuous path such that the end points $T_0$ and $T_1$ have minimal kernel dimension, namely $\dim_\RM(\Ker_\RM(T_0))$ and $\dim_\RM(\Ker_\RM(T_1))$ are either both equal to $0$ or both equal to $1$. The idea in the following is to reduce the definition of the $\ZM_2$-valued spectral flow to the finite dimensional definition, essentially as for the complex spectral flow in \cite{Ph}. There one splits the path into suitably chosen short pieces. We argue analogously. For $a>0$ set
$$
Q_{a}(t)\;=\;
\chi_{(-a,a)}(\imath\,T_t)
\;,
$$
where $\chi_I$ denotes the characteristic function on $I\subset\RM$. This projection is real, that is $\overline{Q_{a}(t)}=Q_{a}(t)$, and is of finite dimensional range for $a$ sufficiently small. Associated to the projections one has the restrictions $Q_{a}(t)\,T_t\,Q_{a}(t)$ which are viewed as skew-adjoint operators on $\Ee_a(t)=\Ran(Q_{a}(t))$. These operators do not have necessarily minimal kernel dimension. This is enforced by adding a skew-adjoint perturbation $R_t$ on the kernel of $Q_{a}(t)\,T_t\,Q_{a}(t)$. The choice of $R_t$ is not necessarily continuous in $t$, as also the dimension of the kernel varies non-continuously with $t$. Now we introduce the following skew-adjoint operators on $\Ee_a(t)$ with minimal kernel dimension:
\begin{equation}
\label{eq-Tachoice}
T^{(a)}_t\;=\;Q_{a}(t)\,T_t\,Q_{a}(t)\,+\,R_t
\;.
\end{equation}
By compactness, it is possible to choose a finite partition $0=t_0<t_1<\ldots<t_{N-1}<t_N=1$ of $[0,1]$ and $a_n> 0$, $n=1,\ldots,N$, such that $t\in [t_{n-1},t_n]\mapsto Q_{a_n}(t)$ is continuous and hence with constant finite rank, and, moreover, for some $\epsilon\leq\tfrac{1}{5}$
\begin{equation}
\label{eq-Qcont}
\|Q_{a_n}(t)-Q_{a_n}(t')\|\;<\;\epsilon
\;,
\qquad
\forall\;\;t,t'\in[t_{n-1},t_n]
\;,
\end{equation}
as well as
\begin{equation}
\label{eq-piTcont}
\|\pi(T_t)-\pi(T_{t'})\|\;<\;\epsilon
\;,
\qquad
\forall\;\;t,t'\in[t_{n-1},t_n]
\;.
\end{equation}
This is illustrated in Figure~\ref{fig-schematic}. Let $\isom_n:\Ee_{a_n}(t_{n-1})\to \Ee_{a_n}(t_{n})$ be the orthogonal projection of $\Ee_{a_n}(t_{n-1})$ onto $\Ee_{a_n}(t_{n})$, namely $\isom_nv=Q_{a_n}(t_{n})v$. By Proposition~\ref{prop-Z2ChangeSpace}, $\isom_n$ is a bijection.

\begin{definition}
\label{def-Z2flow}
For a path $t\in [0,1]\mapsto T_t\in \Ff^1(\Hh_\RM)$ with end points having minimal kernel dimension, let $t_n$ and $a_n$ as well as $T^{(a)}_t$ and $\isom_n$ be as above. Then the  $\ZM_2$-valued spectral flow is defined by 
\begin{equation}
\label{eq-SFdef}
\SF_2(t\in[0,1]\mapsto T_t)
\;=\;
\sum_{n=1}^N 
\SF_2\big(T^{(a_n)}_{t_{n-1}},\isom_n^*T^{(a_n)}_{t_n}\isom_n\big)
\;,
\end{equation}
where on the r.h.s. the $\SF_2$ is the finite dimensional $\ZM_2$-valued spectral flow on $\Ee_{a_n}(t_{n-1})$ defined previously, and the addition is modulo $2$ in $(\ZM_2,+)$.
\end{definition}

The basic result on the $\ZM_2$-valued spectral flow is that it is well-defined by the above procedure. 

\begin{theorem}
\label{theo-welldef} 
Let $t\in[0,1]\mapsto T_t\in\Ff^1(\Hh_\RM)$ be a path with end points having minimal kernel dimension. The definition of $\SF_2(t\in[0,1]\mapsto T_t)$ is independent of the choice of the partition $0=t_0<t_1<\ldots<t_{N-1}<t_N=1$ of $[0,1]$ and the values $a_n>0 $ such that $t\in [t_{n-1},t_n]\mapsto Q_{a_n}(t)$ is continuous and satisfies \eqref{eq-Qcont}, and also the choice of the $R_t$ in \eqref{eq-Tachoice}. Moreover, $\SF_2$ satisfies the concatenation property with a second path $t\in[1,2]\mapsto T_t\in\Ff^1(\Hh_\RM)$
$$
\SF_2(t\in[0,1]\mapsto T_t)
\;+\;
\SF_2(t\in[1,2]\mapsto T_{t})
\;=\;
\SF_2(t\in[0,2]\mapsto T_{t})
\;,
$$ 
with addition in $(\ZM_2,+)$, and is independent of the orientation of the path
$$
\SF_2(t\in[0,1]\mapsto T_t)
\;=\;
\SF_2(t\in[0,1]\mapsto T_{1-t})
\;=\;
\SF_2(t\in[0,1]\mapsto -T_{t})
\;.
$$ 
\end{theorem}

\noindent {\bf Proof.} 
First let us show that adding $R_t$ in \eqref{eq-Tachoice} does not lead to an arbitrariness in the definition of $\SF_2$. Indeed, suppose that $R=R_{t_n}$ is modified to $R'$. This changes two contributions on the r.h.s. of \eqref{eq-SFdef}, namely
$$
I
\;=\;
\SF_2\big(T^{(a_n)}_{t_{n-1}},\isom_n^*T^{(a_n)}_{t_n}\isom_n\big)
\;+\;
\SF_2\big(T^{(a_{n+1})}_{t_{n}},\isom_{n+1}^*T^{(a_{n+1})}_{t_{n+1}}\isom_{n+1}\big)
\;.
$$
Using Proposition~\ref{prop-Z2ChangeSpace} and unraveling the definitions one gets
\begin{align*}
I  \;=\; &
\SF_2\big(
\isom_n T^{(a_n)}_{t_{n-1}}
\isom_n^*,Q_{a_n}(t_{n})\,T_{t_{n}}\,Q_{a_n}(t_{n})\,+\,R{\big)}
\\
& \;+\;
\SF_2\big(
Q_{a_{n+1}}(t_{n})\,T_{t_{n}}\,Q_{a_{n+1}}(t_{n})\,+\,R,\isom_{n+1}^*T^{(a_{n+1})}_{t_{n+1}}\isom_{n+1}\big)
\;.
\end{align*}
Now suppose, say, that $a_{n+1}>a_n$, and set $T_n=Q_{a_{n}}(t_{n})\,T_{t_{n}}\,Q_{a_{n}}(t_{n})$. Then there is some finite dimensional invertible real skewadjoint $T'_n$ such that 
$$
Q_{a_{n+1}}(t_{n})\,T_{t_{n}}\,Q_{a_{n+1}}(t_{n})\,+\,R
\;=\;
T_n\oplus T'_n\,+\,R
\;=\;
(T_n+R)
\oplus T'_n
 \;,
 $$
where it was used that $R$ only acts non-trivially on the kernel of $T_n$. Hence
$$
I
\;=\;
\SF_2\big(
\isom_n T^{(a_n)}_{t_{n-1}}
\isom_n^*,T_n+R\big)
 \;+\;
\SF_2\big(
(T_n+R)
\oplus T'_n,\isom_{n+1}^*T^{(a_{n+1})}_{t_{n+1}}\isom_{n+1}\big)
\;.
$$
Now one has mod $2$
\begin{align*}
0
&
\;=\;
\SF_2\big(T_n+R,T_n+R'\big)
\,+\,
\SF_2\big(T_n+R,T_n+R'\big)
\,+\,
\SF_2\big(T'_n,T'_n\big)
\\
&
\;=\;
\SF_2\big(T_n+R,T_n+R'\big)
\,+\,
\SF_2\big((T_n+R)\oplus T'_n,(T_n+R')\oplus T'_n\big)
\\
&
\;=\;
\SF_2\big(T_n+R,T_n+R'\big)
\,+\,
\SF_2\big((T_n+R')\oplus T'_n,(T_n+R)\oplus T'_n\big)
\;,
\end{align*}
where we appealed to Lemma~\ref{lemma-SF2basic}. Adding this to $I$ and using Proposition~\ref{prop-Z2concat} shows
$$
I
\;=\;
\SF_2\big(
\isom_n T^{(a_n)}_{t_{n-1}}
\isom_n^*,T_n+R'\big)
 \;+\;
\SF_2\big(
(T_n+R')
\oplus T'_n,\isom_{n+1}^*T^{(a_{n+1})}_{t_{n+1}}\isom_{n+1}\big)
\;,
$$
namely the desired independence on the choice of $R$. The remainder of the argument transposes \cite{Ph} to the $\ZM_2$-case, notably we check that the $\ZM_2$-valued spectral flow remains unchanged under (1) refining the partition using the same $a_n$ and (2) keeping the same partition, but changing the $a_n$.  As to (1), let $t_n'\in(t_{n-1},t_n)$ be added to the partition and let the associated value be $a'_n=a_n$. Then there are isomorphisms $\isom'_n:\Ee_{a_n}(t_{n-1})\to \Ee_{a_n}(t'_{n})$ and $\isom''_n:\Ee_{a_n}(t'_{n-1})\to \Ee_{a_n}(t_n)$  defined via orthogonal projections as above. Then we claim that
$$
\SF_2\big(T^{(a_n)}_{t_{n-1}},\isom_n^*T^{(a_n)}_{t_n}\isom_n\big)
\;=\;
\SF_2\big(T^{(a_n)}_{t_{n-1}},(\isom'_n)^*T^{(a_n)}_{t'_n}\isom'_n\big)
\,+\,
\SF_2\big(T^{(a_n)}_{t'_{n}},(\isom''_n)^*T^{(a_n)}_{t_n}\isom''_n\big)
\;,
$$
with addition in $\ZM_2$. This actually follows from Propositions~\ref{prop-Z2ChangeSpace} and \ref{prop-Z2ChangeSpaceCriteria}. For (2), let us suppose that there are $a'_n>a_n$ both satisfying {\eqref{eq-Qcont}}. In particular, both $Q_{a_n}(t)$ and $Q_{a'_n}(t)$ have constant dimension, and thus also the projection $Q_{a'_n}(t)-Q_{a_n}(t)$ has constant dimension. This implies that the added eigenvalues remain in $(a_n,a'_n)$ and $(-a'_n,a_n)$ and thus do not contribute to the $\ZM_2$-valued spectral flow by combining the additivity of Lemma~\ref{lemma-SF2basic}.  Finally, according to Lemma~\ref{lemma-SF2basic},
$$
\SF_2\big(T^{(a_n)}_{t_{n-1}},\isom_n^*T^{(a_n)}_{t_n}\isom_n\big)
\;=\;
\SF_2\big(\isom_nT^{(a_n)}_{t_{n-1}}\isom_n^*,T^{(a_n)}_{t_n}\big)
\;=\;
\SF_2\big(T^{(a_n)}_{t_{n}},\isom_nT^{(a_n)}_{t_{n-1}}\isom_n^*\big)
\;.
$$
This implies the last claim.
\hfill $\Box$

\vspace{.2cm}

The proof of the following result is exactly as that of Proposition~3 in \cite{Ph} and Proposition~2.5 in \cite{Ph1}, provided the local concatenation of spectral flow (based on Lemma 1.3 in \cite{Ph1}) is replaced by Proposition~\ref{prop-Z2ChangeSpace}.

\begin{theorem}
\label{theo-hmot_inv} 
Let $t\in [0,1]\mapsto T_t$ and $t\in[0,1]\mapsto T'_t$ be two continuous paths in $\Ff^1(\Hh_\RM)$ such that $T_0=T'_0$ and $T_1=T'_1$ have minimal kernel dimension. If the two paths are connected via a continuous homotopy leaving the endpoints fixed, $\SF_2(t\in [0,1]\mapsto T_t)=\SF_2(t\in [0,1]\mapsto T'_t)$. 
\end{theorem}

At this point, one may be tempted to write simply $\SF_2(T_0,T_1)$ for $\SF_2(t\in [0,1]\mapsto T_t)$. This is, however, not possible because the $\ZM_2$-valued spectral flow truly depends on the path, and not only on the end points. Indeed, there exist non-trivial loops based at $T_1$ which when concatenated with $t\in [0,1]\mapsto T_t$ change the value of the  $\ZM_2$-valued spectral flow. However, in the case where $\Hh_\RM$ is finite dimensional and $t\in [0,1]\mapsto T_t$ is the linear path (or a homotopy of it), one has $\SF_2(t\in[0,1]\mapsto T_t)=\SF_2(T_0,T_1)$. 

\vspace{.2cm}

As a final issue, let us consider (real) analytic paths in $\Ff^1(\Hh_\RM)$. By analytic perturbation theory \cite[VII.3]{Kat}, all eigenvalues and eigenvectors can be chosen to be analytic. In particular, $T_t$ has minimal kernel dimension except on a discrete set of crossings. At each of these crossings, one {generically} has blocks as in the first example of \eqref{eq-2x2example}, and not the second. Hence each crossing (of simple multiplicity) contributes a unit to the $\ZM_2$-valued spectral flow{. There can be non-generic crossings with a prefactor $(2t-1)^k$ in the left equation of  \eqref{eq-2x2example}. For even $k$, there is vanishing $\ZM_2$-spectral flow, while for odd $k$ it is equal to $1$. Hence }we can conclude the following:

\begin{theorem}
\label{theo-analytic} 
Let $t\in [0,1]\mapsto T_t\in\Ff^1(\Hh_\RM)$ be analytic and have end points with minimal kernel dimension. Then  $\SF_2(t\in [0,1]\mapsto T_t)$ is modulo $2$ equal to the sum of all eigenvalue crossings through $0$ along the path, each one counted with its multiplicity.
\end{theorem}

\section{Index map on the orthogonal group}
\label{sec-prelimOrtho}

The aim of the following two sections is to give an alternative description of the $\ZM_2$-valued spectral flow, that is to express it in terms of the index map on the orthogonal group introduced in \cite{CO}. This section reviews the construction and main properties of this index map from \cite{CO}, but is kept self-contained with all proofs given in a slightly generalized form which will be used in Section~\ref{sec-AlternativeForm} to establish the alternative description. Recall that the orthogonal group on a real Hilbert space is defined as
$$
\Oo(\Hh_\RM)
\;=\;
\left\{O\in\Bb(\Hh_\RM)\;|\;O^*O={OO^*=}\one\right\}
\;.
$$
If the orthogonal operators are viewed as operators on the complexification $\Hh_\CM$, then $\Oo(\Hh_\RM)$ can be identified with the real unitaries:
$$
\Oo(\Hh_\RM)
\;\cong\;
\left\{O\in\Bb(\Hh_\CM)\;|\;O^*O={OO^*=}\one\mbox{ and }\overline{O}=O\right\}
\;.
$$
Let us first suppose that $\dim_\RM(\Hh_\RM)<\infty$. Then one of the basic facts is that $\Oo(\Hh_\RM)$ has two connected components which can be distinguished by the map $j:(\Oo(\Hh_\RM),\cdot)\to(\ZM_2,\cdot)$ defined by
$$
j(O)
\;=\;
\sgn(\det(O))
\;,
$$
where $\det(\one)=1$. This is clearly a homomorphism. Let us suppose that $\dim_\RM(\Hh_\RM)$ is even and $J$ is a given fixed complex structure on $\Hh_\RM$. The one can rewrite $j$ as
\begin{align}
j(O)
& \;=\; 
\tfrac{1}{2}\,\dim_\RM(\Ker_\RM(J+OJO^*))
\;\mbox{\rm mod}\;2
\label{eq-jdef}
\\
&
\;=\; 
\tfrac{1}{2}\,\dim_\RM\big(\Ker_\RM(O- JOJ )\big) \mbox{ mod }  2
\label{eq-kerneq}
\\
&
\;=\; 
\tfrac{1}{2}\,\dim_\RM\big(\Ker_\RM(\one- \tfrac{1}{2}O^*J[O,J] )\big) \mbox{ mod }  2
\;,
\label{eq-kerneqbis}
\end{align}
where in \eqref{eq-jdef} the two representations of $\ZM_2$ are identified, as described after Definition~\ref{def-SF2finite}. To verify \eqref{eq-jdef}, let us note that $\sgn(\det(O))$ is homotopy invariant and so is the r.h.s. of \eqref{eq-jdef} (see \cite{CO} or Proposition~\ref{prop-continuity} below), and the equality can readily be checked to hold for two points in the two components. Let us also point out that \eqref{eq-jdef} is independent of the choice of $J$. The other equalities \eqref{eq-kerneq} and \eqref{eq-kerneqbis} then readily follow. If the dimension of $\Hh_\RM$ is odd, then there is no complex structure and \eqref{eq-jdef} does not hold.

\vspace{.2cm}

In infinite dimension, it is known from Kuipers' theorem that $\Oo(\Hh_\RM)$ is contractible. On the other hand, \eqref{eq-kerneqbis} suggests that an invariant can be defined whenever the commutator $[O,J]$ is compact for a given fixed complex structure $J$ (which always exist on an infinite dimensional Hilbert space). Hence let us set, as in \cite{CO},
\begin{equation}
\label{eq-OJdef}
\Oo_J(\Hh_\RM)
\;=\;
\left\{O\in\Oo(\Hh_\RM)\;|\;[O,J]\in\Kk(\Hh_\RM)\right\}
\;.
\end{equation}
This is actually a subgroup which, as we shall see shortly, is not connected any more. As the orthogonal group acts transitively on the set of complex structures, the subgroups associated to different $J$ are isomorphic. More precisely, if $J'=WJW^*$ for some $W\in \Oo(\Hh_\RM)$, one has $\Oo_{J'}(\Hh_\RM)=W\Oo_J(\Hh_\RM)W^*$. 

\begin{theorem} {\rm \cite{CO}}
\label{theo-jproperties}
The map $j:(\Oo_J(\Hh_\RM),\cdot)\to(\ZM_2,\cdot)$ is a homotopy invariant homomorphism and labels the two connected components of $\Oo_J(\Hh_\RM)$.
\end{theorem}

Furthermore, it is proved in \cite{CO} that $\Oo_J(\Hh_\RM)$ is of the same homotopy type as the loop space of $\Ff^1(\Hh_\RM)$ (this is also shown in the proof of Theorem~\ref{theo-isomorphism} below). Theorem~\ref{theo-jproperties} follows from the following more general continuity result upon setting $J_1=J$ and $J_2=OJO^*$.

\begin{prop}
\label{prop-continuity}
Introduce the set
$$
\Pp
\;=\;
\left\{(J_0,J_1)\in \Bb(\Hh_\RM)\times \Bb(\Hh_\RM)\,\left|\,
J_i^2=-\one\;\mbox{\rm and }
J_i^*=-J_i\;\mbox{\rm for }i=0,1\,,\; ||J_1-J_0||_{\mathcal Q}<\tfrac{1}{2}
\right.\right\}
\;,
$$
equipped with the norm topology. Then
\begin{equation}
\label{eq-Z2def}
(J_0,J_1)\,\in\,\Pp
\;\mapsto\;
\left(
\tfrac{1}{2}\,\dim_\RM(\Ker_\RM(J_0+J_1))\right)\;\mbox{\rm mod}\; 2
\;\in\;\ZM_2
\;,
\end{equation}
is continuous. 
\end{prop}

The proof rests on the following lemma.

\begin{lemma}\label{ident} With the assumptions of Proposition~\ref{prop-continuity} on the pair $J_0,J_1\in\mathcal P$,
let 
$$
T_0\;=\;\tfrac{1}{2}\,(J_0+J_1)
\;,
\qquad
T_1\;=\;\tfrac{1}{2}\,(J_0-J_1)
\;.
$$
Then the following identities hold:
$$
T_0^*T_0+T_1^*T_1\;=\;\one\;=\;T_0T_0^*+T_1T_1^*\;,
\qquad
T_0^*T_1+T_1^*T_0\;=\;0\;=\;T_0T_1^*+T_1T_0^*\;,
$$
as well as
$$
T_0J_0\;=\;J_1T_0\;,
\qquad
T_0J_1\;=\;J_0T_0\;,
\qquad
T_1J_0\;=\;-J_1T_1\;,
\qquad
T_1J_1\;=\;-J_0T_1\;.
$$
\end{lemma}

{\bf Proof.}
The arguments are purely algebraic. Let us start from
$$
T_0^*T_0
\;=\;
-T_0^2
\;=\;
-\tfrac{1}{4}(J_0^2+J_1^2+J_0J_1+J_1J_0)
\;=\;
\tfrac{1}{2}-\tfrac{1}{4}(J_0J_1 +J_1J_0)
\;,
$$
as well as
$$
T_1^*T_1
\;=\;
-T_1^2
\;=\;
\tfrac{1}{2}\,+\,\tfrac{1}{4}(J_1J_0+J_0J_1)
\;.
$$
The first identity now follows and second one uses the same algebraic relations. The final group of identities are all proved in the same way, for example: $T_0J_0=\frac{1}{2}(-\one+J_1J_0)$ while $J_1T_0 =\frac{1}{2} (J_1J_0-\one)$ which gives the first identity.
\hfill $\Box$

\vspace{.2cm}

{\bf Proof} of Proposition~\ref{prop-continuity}.
By the assumption on the norm of the difference $J_1- J_0$ in the Calkin algebra, $T_0$ is a skew-adjoint Fredholm. The focus is on $\Ker(T_0)=\Ker(T_0^*T_0)$. First of all, the identities in Lemma \ref{ident} imply
$$
T_0^*T_0J_0
\;=\;
-T_0(T_0J_0)
\;=\;
-T_0(J_1T_0)
\;=\;
-(T_0J_1)T_0
\;=\;
-(J_0T_0)T_0
\;=\;
J_0(T_0^*T_0)
\;.
$$
Hence $T_0^*T_0$ is a complex linear operator on $\Hh_\RM$ equipped with $J_0$ as a complex structure. This implies that $\Ker(T_0)=\Ker(T_0^*T_0)$ is always even dimensional as a real vector space. It follows that the map in \eqref{eq-Z2def} is well-defined and really takes values in $\ZM_2=\{0,1\}$. 

\vspace{.2cm}

We now claim that all eigenvalues $\lambda\in(0,1)$ of the non-negative Fredholm operator $T_0^*T_0$ have even complex multiplicity, which implies that their real multiplicity is divisible by $4$. To see this, let a non-vanishing $v\in\Hh_\RM$ be such that $T_0^*T_0v=\lambda v$ with $\lambda\in(0,1)$. Then $J_0v$, its multiple by the imaginary unit, is also an eigenvector of $T_0^*T_0$ with eigenvalue $\lambda$. It is linearly independent of $v$ over the reals, but in the complex Hilbert space it is, of course, linearly dependent on $v$. Moreover, the relations of Lemma \ref{ident} show that $w=T_1^*T_0v$ is also an eigenvector of $T_0^*T_0$:
$$
T_0^*T_0w
\;=\;
T_0^*T_0T_1^*T_0v
\;=\;
-T_0^*T_1T_0^*T_0v
\;=\;
T_1^*T_0T_0^*T_0v
\;=\;
T_1^*T_0\,\lambda v
\;=\;
\lambda w
\;.
$$
The norm of this vector is given by $\|w\|^2=v^*T_0^*(1-T_0T_0^*)T_0v=\lambda(1-\lambda)\|v\|^2$ so that it is non-vanishing for $\lambda\not=0,1$, and furthermore $w$ is linearly independent of $v$ as complex vector. Indeed, suppose that $w=(\mu_0+\mu_1 J_0)v$ with some $\mu_0,\mu_1\in\RM$ not both zero. Then applying $T_0^*T_1$ yields, again using the relations of Lemma~\ref{ident},
\begin{align*}
T_0^*T_1T_1^*T_0v= T_0^*T_1(\mu_0+\mu_1 J_0)v
\quad
& \Longrightarrow
\quad
T_0^*T_0(\one-T_0^*T_0)v=-(\mu_0-\mu_1 J_0) T_1^*T_0v
\\
& 
\Longrightarrow
\quad
\lambda(1-\lambda)v=-(\mu_0-\mu_1 J_0) w
\\
& 
\Longrightarrow
\quad
\lambda(1-\lambda)(\mu_0+\mu_1 J_0)v=-(\mu_0^2+\mu_1^2) w
\;,
\end{align*}
where in the last implication we applied $\mu_0+\mu_1 J_0$. Hence $\lambda(1-\lambda)w=-(\mu_0^2+\mu_1^2) w$ which is a contradiction because $\lambda\leq 1$. Thus $v,w$ span a two dimensional complex Hilbert space of eigenvectors for $T_0^*T_0$
with eigenvalue $\lambda$.

\vspace{.2cm}

Suppose that $u$ is another eigenvector of $T_0^*T_0$ with eigenvalue $\lambda$ that is orthogonal to the real span $\Ee$ of $\{v, J_0v, T_1^*T_0v, J_0T_1^*T_0v\}$. Then the span of  $\{u, J_0u, T_1^*T_0u, J_0T_1^*T_0u\}$ can seen to be orthogonal to $\Ee$ showing that each eigenspace of $T_0^*T_0$ is a direct sum of these four (real) dimensional subspaces. Given that the degeneracy of every positive eigenvalue of $T_0^*T_0$ is divisible by $4$, the result now follows. (Note that this argument is partly modeled on that in Proposition 5.1 of \cite{AS}.)
\hfill $\Box$

\vspace{.2cm}

\noindent {\bf Remark} Let us stress that the above also proves the following somewhat surprising fact from linear algebra. For two complex structures $J_0$ and $J_1$ on $\RM^{2n}$, the multiplicity of every eigenvalue of $(J_0+J_1)^2$ in $(-1,0)$ is divisible by $4$. 
\hfill $\diamond$

\vspace{.2cm}

In the remainder of this section, we provide an alternative formula for the map $j$.

\begin{prop}
\label{prop-contract} Every $O\in\Oo_J(\Hh_\RM)$ can be written as $O=U(\one+K)$ with an orthogonal $U\in\Oo(\Hh_\RM)$ satisfying $JU=UJ$ and a compact operator $K\in\Kk(\Hh_\RM)$. One has
\begin{equation}
\label{eq-jalt}
j(O)\;=\;\dim_\RM(\Ker_\RM(K+2\,\one))\;\mbox{\rm mod }  2
\;.
\end{equation}
Moreover, $\Oo_J(\Hh_\RM)$ can be retracted to the subgroup $\Oo_\Kk(\Hh_\RM)=\left\{O\in\Oo(\Hh_\RM)\;|\;O-\one\in\Kk(\Hh_\RM)\right\}$.
\end{prop}

\noindent {\bf Proof.} The first claim is Proposition~2.1 in \cite{CO}, but we here provide an explicit formula for $U$. Set $S_0=\frac{1}{2}(O-JOJ)$ and $S_1=\frac{1}{2}(O+JOJ)$ so that $O=S_0+S_1$. The formulas from Lemma~\ref{ident} will be used for $J_1=J$ and $J_2=O^*JO$. As then $S_0=OT_0J^*$ and $S_1=OT_1J^*$, one has 
$$
S_0^*S_0+S_1^*S_1\;=\;\one\;=\;S_0S_0^*+S_1S_1^*\;,
\qquad
S_0J\;=\;JS_0\;,
\qquad
S_1J\;=\;-J\,S_1
\;.
$$
Let us use the polar decomposition $S_0=V|S_0|$. Then $VJ=JV$, but, in general, $V$ is only a partial isometry with kernel $\Ker_\RM(S_0)=\Ker_\RM(|S_0|)$. This kernel is $J$-invariant and real, and therefore even dimensional. Let us choose an operator $I$ on $\Ker_\RM(S_0)$ with $I^2=\one$ and $IJ=-JI$. Then the multiplicities of $1$ and $-1$ as eigenvalues of $I$ are equal. Now define $U$ as $V$ on $\Ker_\RM(S_0)^\perp$ and $S_1I$ on $\Ker_\RM(S_0)$. Then $U$ is orthogonal and satisfies $UJ=JU$. Furthermore, $O=U(|S_0|+U^*S_1)$ so that $K=|S_0|-\one+U^*S_1$, which is indeed compact because $S_1=\frac{1}{2}J[O,J]$ is compact. Moreover, $U^*S_1=I$ on $\Ker_\RM(S_0)$. Thus the multiplicity of $-1$ as eigenvalue of $\one+K$ is equal to $\frac{1}{2}\dim_\RM(\Ker_\RM(S_0))$. Comparing with \eqref{eq-jdef}, this shows the formula for $j$. The final claim follows directly from Kuipers' theorem because $U$ is unitary on $\Hh_\RM$ viewed as complex Hilbert space with imaginary unit $J$.
\hfill $\Box$

\vspace{.2cm}

\noindent {\bf Example} Given a one-dimensional projection $P$ on $\Hh_\RM$, let us set $O=\one-2P$. Then $O\in \Oo_\Kk(\Hh_\RM)$ and $j(O)=1$ by \eqref{eq-jalt}. If, moreover, $PJP=0$ holds, one can readily check the identity $\one- \tfrac{1}{2}O^*J[O,J]=\one-2P$ so that also \eqref{eq-kerneqbis} leads to $j(O)=1$.
\hfill $\diamond$

\vspace{.2cm}

\noindent {\bf Remark} An alternative proof of \eqref{eq-jalt} can be given as follows. The r.h.s. of \eqref{eq-jalt} is a homotopy invariant because the spectrum of every orthogonal $\one+K$ is invariant under complex conjugation so that the parity of the $-1$ eigenvalue is conserved.  By Theorem~\ref{theo-jproperties} also $j$ is a homotopy invariant. Hence it is sufficient to check the equality on the two components. This is trivial for the identity component and was verified on the other component in the example above.
\hfill $\diamond$

\section{Alternative formulation of the $\ZM_2$-valued spectral flow}
\label{sec-AlternativeForm}

Let us begin by considering the straight-line path connecting two complex structure on a real Hilbert space. The following lemma on the spectral properties along this path is elementary.

\begin{lemma}
\label{lem-basics}
Let $J_0$ and $J_1$ be complex structures on $\Hh_\RM$ such that $\|\pi(J_0)-\pi(J_1)\|_{\Qq}\leq c<1$. Set $T_t=(1-t)J_0+tJ_1$ for $t\in[0,1]$. Then $t\in[0,1]\mapsto T_t$ is a path in $\Ff^1(\Hh_\RM)$ such that $\sigma_\ess(T_t)\cap [-\imath(1-ct),\imath(1-ct)]=\emptyset$. Furthermore, $0\in\sigma(T_t)$ implies that $t=\frac{1}{2}$.
\end{lemma}

Given the situation of Lemma~\ref{lem-basics}, $t\in[0,1]\mapsto \imath T_t$ is a path of self-adjoint Fredholms and it is hence possible to consider the associated spectral flow $\SF(t\in[0,1]\mapsto \imath T_t)$ in the sense of \cite{Ph1}. Note that as this path is analytic in $t$, analytic perturbation theory for the discrete spectrum of the self-adjoints $\imath T_t$ on $\Hh_\CM$ applies so that all notions of spectral flow coincide. In particular, all eigenvalues crossings through $0$ at $t=\frac{1}{2}$ result from analytic curves. Hence the spectral symmetry $\sigma(\imath T_t)=-\sigma(\imath T_t)$ implies that each analytic curve of an eigenvalue has a reflected partner, and the total spectral flow resulting from each such a pair vanishes. In conclusion, $\SF(t\in[0,1]\mapsto \imath T_t)=0$. Now the $\ZM_2$-valued spectral flow counts the number of eigenvalue exchanges at $t=\frac{1}{2}$. An important point is that the present path is analytic, hence locally at the crossing the first example in \eqref{eq-2x2example} is a good model, while the second is {\em not}. Therefore the number of crossings and thus the $\ZM_2$-valued spectral flow is given by the r.h.s. of \eqref{eq-reformulate}. This will be shown in more detail below. As is already hinted at in Lemma~\ref{lem-basics}, the number of these crossings can be read off from the kernel dimension of $T_{\frac{1}{2}}=\frac{1}{2}(J_0+J_1)$.  Before stating this result, let us introduce the following 

\vspace{.2cm}

{\bf Notation:} {\it $\SF_2(J_0,J_1)$ denotes $\SF_2(t\in[0,1]\mapsto T_t)$ for the linear path described in Lemma~\ref{lem-basics}.}

\begin{prop}
\label{prop-basicSF}
Let $J_0$ and $J_1$ be complex structures on $\Hh_\RM$ such that $\|\pi(J_0)-\pi(J_1)\|_{\Qq}\leq c<1$. Then the $\ZM_2$-valued spectral flow from $J_0$ to $J_1$ is 
\begin{equation}
\label{eq-reformulate}
\SF_2(J_0,J_1)
\;=\;
\big(\tfrac{1}{2}\,\dim_\RM(\Ker_\RM(J_0+J_1))\big)\;\mbox{\rm mod}\;2
\;.
\end{equation}
\end{prop}

Note that as $J_0+J_1$ is in the component of $\Ff^1(\Hh_\RM)$ with vanishing $\ZM_2$-index, the kernel of $J_0+J_1$ is always even dimensional (as real vector space if $J_0+J_1$ is viewed as an $\RM$-linear operator, and as a complex vector space is $J_0+J_1$ is viewed as an $\CM$-linear operator). Hence the r.h.s. of \eqref{eq-reformulate} is indeed well-defined in $\ZM_2$. 

\vspace{.2cm}

{\bf Proof} of Proposition~\ref{prop-basicSF}. 
Let us calculate $\SF_2(J_0,J_1)$ as given by definition \eqref{eq-SFdef} for the special straight line path of Lemma~\ref{lem-basics}. It is possible to choose a splitting $t_0=0<t_1<\frac{1}{2}<t_2=1-t_1<t_3=1$ of $[0,1]$ in three intervals as well as $a>b$ with the following properties (see one of the crossings in Figure~\ref{fig-schematic}): $Q_{b}(t)=0$ for $t\in [t_0,t_1]$ and $Q_{b}(t)=0$ for $t\in [t_2,t_3]$, and $\Tr(Q_{a}(t))=\dim_\RM(\Ker_\RM(T_{\frac{1}{2}}))$ for $t\in[t_1,t_2]$. Note that, in particular, $\sigma(T^{(a)}_{\frac{1}{2}})=\{0\}$. Only the interval $[t_1,t_2]=[t_1,1-t_1]$ contributes to the $\ZM_2$-valued spectral flow. Hence, with the notations of Section~\ref{sec-def},
$$
\SF_2(J_0,J_1)
\;=\;
\SF_2\big(T^{(a)}_{t_{1}},\isom_2^*T^{(a)}_{t_2}\isom_2\big)
\;.
$$
Now both finite dimensional skew-adjoint operators $T^{(a)}_{t_{1}}$ and $\isom_2^*T^{(a)}_{t_2}\isom_2$ are non-degenerate. Using the polar decomposition, each operator $T^{(a)}_t$ with $t\not=\frac{1}{2}$ can hence be homotopically deformed to complex structures $J^{(a)}_t$. Actually, if $J_t$ is the (skew-adjoint) phase of $T_t$ for $t\not=\frac{1}{2}$, then $J^{(a)}_t=Q_a(t)J_t Q_a(t)$. As there is no kernel along both of these homotopies,
$$
\SF_2(J_0,J_1)
\;=\;
\SF_2(J^{(a)}_{t_1},\isom_2^*J^{(a)}_{t_2}\isom_2)
\;.
$$
As the orthogonal group acts transitively on complex structures, there exists an orthogonal $O$ such that $J^{(a)}_{t_2}=O^*J^{(a)}_{t_1} O$. From the definition of $\SF_2$ and \eqref{eq-jdef}, one now has
$$
\SF_2(J_0,J_1)
\;=\;
\sgn(\det(O))
\;=\;
\tfrac{1}{2}\,\dim_\RM(\Ker_\RM(J^{(a)}_{t_1}+\isom_2^*J^{(a)}_{t_2}\isom_2))
\;\mbox{\rm mod}\;2
\;.
$$
On the other hand, $J_0$ is homotopic to $J_{t_1}$ and $J_1$ is homotopic to $J_{t_2}$. Thus by Proposition~\ref{prop-continuity}
$$
\tfrac{1}{2}\,\dim_\RM(\Ker_\RM(J_0+J_1)
\;\mbox{\rm mod}\;2
\;=\;
\tfrac{1}{2}\,\dim_\RM(\Ker_\RM(J_{t_1}+J_{t_2}))
\;\mbox{\rm mod}\;2
\;.
$$
Moreover, $T_{t_1}$ and $T_{t_2}$ can be made arbitrarily close by sending $t_1$ to $\frac{1}{2}$ (but {\em not} $J_{t_1}$ and $J_{t_2}$). Due to the continuity of the associated Riesz projections, $\isom_2$ can hence be extended to an invertible operator on all $\Hh_\RM$ which is close to the identity and satisfies $\one-J^{(a)}_{t_1}=\isom_2^*(\one-J^{(a)}_{t_2})\isom_2$. Again using the homotopy invariance of Proposition~\ref{prop-continuity} to deform $\isom_2$ to the identity, one concludes
\begin{align*}
\tfrac{1}{2}\,\dim_\RM(\Ker_\RM(J_0+J_1)
\;\mbox{\rm mod}\;2
& \;=\;
\tfrac{1}{2}\,\dim_\RM(\Ker_\RM(J_{t_1}+\isom_2^* J_{t_2}\isom_2))
\;\mbox{\rm mod}\;2
\\
& \;=\;
\tfrac{1}{2}\,\dim_\RM(\Ker_\RM(J^{(a)}_{t_1}+\isom_2J^{(a)}_{t_2}\isom_2^*))
\;\mbox{\rm mod}\;2
\;,
\end{align*}
where in the second equality we used  $\isom_2^* J_{t_2}\isom_2=\one-J^{(a)}_{t_1}+\isom_2^* J^{(a)}_{t_2}\isom_2$. Combined with the above, this concludes the proof.
\hfill $\Box$

\vspace{.2cm}

Now we can write out the alternative formulation of the $\ZM_2$-valued spectral flow defined in \eqref{eq-SFdef}. For the sake of simplicity, let us restrict to a norm continuous path $t\in [0,1]\mapsto T_t\in\Ff^1(\Hh_\RM)$ in the space of skew-adjoint real Fredholm operators with even-dimensional kernel, namely $\Ind_1(T_t)=0$. The end points $T_0$ and $T_1$ are supposed to have trivial kernel. Associated to $T_t$ is the phase $W_t=T_t|T_t|^{-1}$. One has $(W_t)^*=-W_t$, and $(W_t)^*W_t=-W_t^2$ is the projection onto the orthogonal complement of the kernel of $T_t$. Furthermore, $\pi(W_t)\in\Qq(\Hh_\RM)$ is a complex structure in the Calkin algebra, namely $\pi(W_t)$ is skew-adjoint and squares to minus the identity. The map $t\mapsto \pi(W_t)$ is continuous, but $t\mapsto W_t$ is {\em not}, as it is discontinuous at points where the kernel dimension of $T_t$ changes. In any case, one can complete $W_t$ on its kernel by an arbitrary complex structure to obtain a complex structure $J_t$ on $\Hh_\RM$, similar as in \eqref{eq-Tachoice}. Now let $t_0=0< t_1< \ldots< t_n=1$ be a partition as in  \eqref{eq-SFdef}. Let us further assume the partition to be sufficiently fine so that each pair $J_{t_{n-1}} ,J_{t_{n}}$ satisfies the assumptions of Proposition~\ref{prop-basicSF} (this is possible because the difference $\pi(J_{t_{n-1}} -J_{t_{n}})$ can be written out using Riesz projections and resolvent identity and then estimated using \eqref{eq-piTcont}). Then the concatenation property of $\SF_2$ implies that, with addition modulo $2$ in $(\ZM_2,+)$,
$$
\SF_2(t\in[0,1]\mapsto T_t)
\;=\;
\sum_{n=1}^N  \SF_2(J_{t_{n-1}} ,J_{t_{n}})
\;.
$$
Let us now assume the partition to be sufficiently fine so that each pair $J_{t_{n-1}} ,J_{t_{n}}$ satisfies the assumptions of Proposition~\ref{prop-basicSF}. Then we obtain the following formula, which is the $\ZM_2$-equivalent of the index formulation of the complex spectral flow given in \cite{Ph1,BCPRSW,DS2}.

\begin{prop}
Let $t\in[0,1]\mapsto T_t\in\Ff^1(\Hh_\RM)$ be a continuous path with end points having trival kernel. Let $J_t$ be complex structures obtained by completing the phase $T_t|T_t|^{-1}$ by an arbitrary complex structure on the kernel. Then, for a sufficiently fine partition $t_n$ satisfying $\|\pi(J_n-J_{n-1})\|<1$, one has 
$$
\SF_2(t\in[0,1]\mapsto T_t)
\;=\;
\Big(
\tfrac{1}{2}
\sum_{n=1}^N  \dim_\RM(\Ker_\RM(J_{t_{n-1}} +J_{t_{n}}))
\Big)
\;\mbox{\rm mod}\;2
\;.
$$
\end{prop}

\section{The isomorphism on the fundamental group}

Let us first note that for loops there is no need to impose any conditions on the kernel dimension of the end point. Thus $\SF_2$ is a well-defined map on the set of loops in $\Ff^1({\mathcal H}_{\mathbb R})$.

\begin{theorem} 
\label{theo-isomorphism}
The map $\SF_2$ on loops in $\Ff^1({\mathcal H}_{\mathbb R})$ is a homotopy invariant and induces an isomorphism of $\pi_1(\Ff^1({\mathcal H}_{\mathbb R}))$ with $\ZM_2$.
\end{theorem}

{\bf Proof.} The argument follows closely \cite{AS}, p. 11, and \cite{P2}, as well as Subsection 2.8 of \cite{Ph1}. Let $\rho:\Ff^1(\Hh_\RM)\to\Ff^1(\Hh_\RM)$ be the (non-linear and discontinuous) map sending $T$ to the partial isometry $W=T|T|^{-1}$ in the polar decomposition. If $\pi$ denotes the projection onto the Calkin algebra, then the map $\rho_\Qq=\pi\circ\rho$ sends $\Ff^1(\Hh_\RM)$ surjectively onto  the space of complex structures in the Calkin algebra given by
$$
\Cc(\Hh_\RM)
\;=\;
\{\pi(J)\in\Qq(\Hh_\RM)\,|\,\pi(J)^*=-\pi(J)\,,\;\pi(J)^*\pi(J)=\one\}
\;.
$$
The Bartle-Graves selection theorem (see \cite{BD} for a modern proof) provides a right inverse $\theta:\Cc(\Hh_\RM)\to \Ff^1(\Hh_\RM)$ to $\rho_\Qq$, namely $\rho_\Qq\circ \theta=\one$. Moreover, $\theta\circ \rho_\Qq$ is homotopic to the identity via the homotopy $t\in[0,1]\mapsto t\,T+(1-t)\,\theta(\rho_\Qq(T))\in\Ff^1(\Hh_\RM)$. Thus $\rho_\Qq$ is actually a homotopy equivalence so that, in particular, $\Ff^1(\Hh_\RM)$ and $\Cc(\Hh_\RM)$ have the same homotopy groups.

\vspace{.2cm}

Let us next fix a complex structure $J$ on $\Hh_\RM$ which also specifies a base point $\rho_\Qq(J)$ in $\Cc(\Hh_\RM)$. Associated to $J$, one can define a map $\beta_J:\Oo(\Hh_\RM)\to\Cc(\Hh_\RM)$ via $\beta_J(O)=\rho_\Qq(OJO^*)$. This map is actually a Serre fibration by the argument in Theorem 3.9 of \cite{P2}. The fiber over the base point $\rho_\Qq(J)$ is precisely the set $\Oo_J(\Hh_\RM)$ from \eqref{eq-OJdef}. Hence one can dispose of the long exact sequence of homotopy groups, which due to the triviality of the homotopy groups of $\Oo(\Hh_\RM)$  implies that the set $\Omega_{\rho_\Qq(J)}\Cc(\Hh_\RM)$ of based loops in the base space is homotopy equivalent to the fiber over the base point which here is $\Oo_J(\Hh_\RM)$. Combined with the above, we conclude that the based loop space $\Omega_J\Ff^1(\Hh_\RM)$ is homotopy equivalent to $\Oo_J(\Hh_\RM)$. Since $\pi_0(\Oo_J(\Hh_\RM))=\ZM_2$ and due non-trivial examples (Section~\ref{sec-ex} or Proposition~\ref{prop-fluxflow}), this proves the claim.
\hfill $\Box$

\vspace{.2cm}

As in \cite{Ph1}, one can be more explicit about this map.  A loop $t\in[0,1]\mapsto T_t$ in the skew-adjoint Fredholms based at $T=T_0=T_1$, say with $\rho_\Qq(T)=\rho_\Qq(J)$, pushes down to a based loop in $\Cc(\Hh_\RM)$.  This lifts to a path in $\Oo({\mathcal H}_{\mathbb R})$ with endpoints $\one$ and $O\in\Oo_J(\Hh_\RM)$. Then $\SF_2(t\in[0,1]\mapsto T_t)=j(O)$.

\section{An index formula}
\label{sec-index}

The conventional spectral flow can always be expressed as an (Noether) index of an associated Toeplitz operator, see \cite{Ph,Ph1,BCPRSW,DS2}. The following result is the $\ZM_2$-equivalent of this result.

\begin{theorem} 
\label{theo-index}
Let $J$ be a complex structure and $O$ an orthogonal operator on $\Hh_\RM$ such that $[J,O]$ is compact. Extend $J$ to a skew-adjoint operator on $\Hh_\CM=\Hh_\RM\otimes\CM$ and let $P$ be the spectral projection onto the positive imaginary spectrum. Then
$$
\SF_2(J,OJO^*)
\;=\;
\dim_\CM\big(\Ker_\CM(POP|_{P\Hh_\CM})\big)\;\mbox{\rm mod} \,2
\;.
$$
\end{theorem}

Here the $\ZM_2$-index on the r.h.s. is of the type $(j,d)=(2,8)$ in Theorem~1 of \cite{GS} which is proved in Section~2.2.4 therein. Indeed, $P$ satisfies $\overline{P}=\one-P$ (namely, $P$ is even Lagrangian in the terminology of \cite{GS}) and $\overline{O}=O$ with complex conjugation in $\Hh_\CM$. In particular, the index pairing on the r.h.s. is a homotopy invariant under variations of $O$ and $P$ respecting the two symmetries mentioned above. The proof of Theorem~\ref{theo-index} is remarkably simple. Both sides of the equality are homotopy invariants and lie in $\ZM_2$. Hence it is sufficient to verify equality on each component. For $O=\one$, both sides vanish. For the other component, the equality is verified for a non-trivial example in the next section.

\section{An example of non-trivial $\ZM_2$-valued spectral flow \label{example}}
\label{sec-ex}

This section introduces the real analogue of classical Toeplitz operators on $L^2(\SM^1)$. Furthermore it is an important element of the proof of Theorem~\ref{theo-index}.

\vspace{.2cm}

Let us consider the real Hilbert space $\Hh_\RM=L^2_\RM(\SM^1)\otimes\RM^2$, as well as its complexification $\Hh_\CM=\Hh_\RM\otimes\CM=L^2(\SM^1)\otimes\CM^2$. The complex conjugation on $\Hh_\CM$ is denoted by $\Cc$. Next let us consider the discrete Fourier transform 
$$
\Ff:\ell^2(\ZM)\otimes\CM^2\to
\Hh_\CM\;,
\qquad
(\Ff\phi)(k)
\;=\;
\frac{1}{\sqrt{2\pi}}\,\sum_{n\in\ZM}e^{\imath kn}\phi_n
\;.$$
for $\phi=(\phi_n)_{n\in\ZM}\in\ell^2(\ZM)\otimes\CM^2$ with $\phi_n\in \mathbb C^2$. Denote the natural complex conjugation on $\ell^2(\ZM)\otimes\CM^2$ also by $\Cc$ and   the reflection on $\ell^2(\ZM)\otimes\CM^2$ by $\Rr$ (namely $\Rr\phi=(\phi_{-n})_{n\in\ZM}$), one then has $\Ff\Cc=\Rr\Cc\Ff$. 

\vspace{.2cm}

Now let us introduce an operator $\widehat{J}$ on $\ell^2(\ZM)\otimes\CM^2$ by
\begin{equation}
\label{eq-Jhatex}
\widehat{J}
\;=\;
\imath\,\sgn(X)\otimes\one_2\;+\;p_0\otimes\sigma
\;,
\end{equation}
where $p_n$ is the projection on the $n^{\mbox{\rm\tiny th}}$ component in $\ell^2(\ZM)$ and 
$$
\sgn(X)\;=\;
\sum_{n> 0}p_n\,-\,\sum_{n<0}p_n
$$ 
is the sign of the position operator $X=\sum_n np_n$, and $\sigma=\binom{0\;-1}{1\;\;\;0}$.  All this assures that 
$$
\widehat{J}^{\,2}\;=\;-\,\one
\;,
\qquad
\Rr\,\Cc\,\widehat{J}\,\Cc\,\Rr\;=\;\widehat{J}
\;.
$$
This implies that $J=\Ff^*\widehat{J}\Ff$ on $\Hh_\CM$ satisfies $J^2=-\one$ and $\Cc J\Cc=J$. Therefore $J$ restricts to $\Hh_\RM$ and defines a complex structure there. On basis vectors, it is explicitly given by
$$
J
\begin{pmatrix}
c+\cos(nk)+\sin(mk) \\
c'+\cos(n'k)+\sin(m'k)
\end{pmatrix}
\;=\;
\begin{pmatrix}
-c'-\sgn(n)\cos(nk)+\sgn(m)\sin(mk) \\
c-\sgn(n')\cos(n'k)+\sgn(m')\sin(m'k)
\end{pmatrix}
\;,
$$
where $c,c'\in\RM$ and $n,m,n',m'\in\NM$.

\vspace{.2cm}

Next, another complex structure $OJO^*$ on $\Hh_\RM$ will be constructed by conjugating with an orthogonal $O=(O(k))_{k\in\SM^1}$ which we choose simply to be 
$$
O(k)
\;=\;
\begin{pmatrix} \cos(k) & -\sin(k) \\ \sin(k) & \cos(k) \end{pmatrix}
\;.
$$
Our first aim in the calculation below is to show that the $\ZM_2$-valued spectral flow from $J$ to $OJO^*$ along the straight line path  $t\in[0,1]\mapsto J_t=(1-t)J+t\,OJO^*$ is
\begin{equation}
\label{eq-SFex}
\SF_2(J,OJO^*)\;=\;1
\;.
\end{equation}
Actually, it will become apparent that it arises from a single block of the form $T_t$ given in \eqref{eq-2x2example}. It is now also possible to close the path $J_t$ to a closed loop as follows. By Kuipers' theorem there exists a path $t\in[0,1]\mapsto O_t$ connecting $O_0=O$ to $O_1=\one$. Then set $J_t=O_{t-1}JO_{t-1}^*$ for $t\in[1,2]$. Then $t\in[0,2]\mapsto J_t$ is a closed loop with non-trivial $\ZM_2$-valued spectral flow.

\vspace{.2cm}

The second aim is to show that the associated Toeplitz operator $POP$ has a non-trivial $\ZM_2$-index: 
\begin{equation}
\label{eq-Toepex}
\dim_\CM\big(\Ker_\CM(POP|_{P\Hh_\CM})\big)\,\mbox{\rm mod}\,2
\;=\;1
\;.
\end{equation}
Here $P$ is the spectral projection of $J$ viewed as an operator on $\Hh_\CM$ corresponding to the eigenvalue $\imath$. Hence \eqref{eq-SFex} and \eqref{eq-Toepex} together provide an instance in which Theorem~\ref{theo-index} holds in the non-trivial component.

\vspace{.2cm}

The verification of \eqref{eq-SFex}  is easiest on the Fourier transform. Hence let us begin by noting that
$$
\widehat{O}\;=\;
\Ff\,O\,\Ff^*
\;=\;
\frac{1}{2}
\begin{pmatrix} S+S^* & \imath (S^*-S) \\ \imath (S-S^*) & S+S^*
\end{pmatrix}
\;,
$$
where $S$ is the right bilateral shift operator on $\ell^2(\ZM)$. 
Note that as $O$ is real, one has $\Rr\Cc\widehat{O}\Cc\Rr=\widehat{O}$. Furthermore, $S=\Cc S\Cc=\Rr S^* \Rr$. If now $\pi_n:\CM\to\ell^2(\ZM)$ denotes the partial isometric embedding onto the $n^{\mbox{\rm\tiny th}}$ component (so that $p_n=\pi_n(\pi_n)^*$), one also has
$$
\sgn(X)\,S\;=\;S\,\sgn(X)\,+\,\pi_0 (\pi_{-1})^*\otimes\one_2\,+\,\pi_1(\pi_0)^*\otimes\one_2
\;.
$$
Using this and some care and patience, one can check that the first summand in \eqref{eq-Jhatex} satisfies
$$
\widehat{O}\,(\imath\,\sgn(X)\otimes\one_2)\,\widehat{O}^*
\;=\;
\imath\,\sgn(X)\otimes\one_2\,+\,\tfrac{1}{2}
\big(
p_{-1}\otimes (\imath\one_2-\sigma)
-
2\,p_0\otimes \sigma
+
p_1\otimes (-\imath\one_2-\sigma)
\big)
\;.
$$
Another calculation shows that for the second summand in \eqref{eq-Jhatex} 
$$
\widehat{O}\,(p_0\otimes\sigma)\,\widehat{O}^*
\;=\;
-\,\tfrac{1}{2}
\big(
p_{-1}\otimes (\imath\one_2-\sigma)
+
p_1\otimes (-\imath\one_2-\sigma)
\big)
\;.
$$
Note that all these terms are invariant under conjugation with $\Cc\Rr$, as they should be. Combining, one deduces
$$
\widehat{O}\,\widehat{J}\,\widehat{O}^*
\;=\;
\imath\,\sgn(X)\otimes\one_2\;-\;p_0\otimes\sigma
\;,
$$
and thus as claimed above
$$
\widehat{J}_t
\;=\;
\imath\,\sgn(X)\otimes\one_2
\;+\;
(1-2t)\,
p_0\otimes \sigma
\;.
$$
The example illustrates that the fundamental spectral unit in this game is a copy of $\mathbb R^2$. Given a phase $W_1$ of a skew-adjoint Fredholm, the Hilbert space decomposes as a direct sum of its kernel plus its orthogonal complement. Off the kernel, we can find a basis of the Hilbert space such that the whole Hilbert space  is a direct sum of two dimensional subspaces on which $W_1$ acts as the matrix $\sigma$.

\vspace{.2cm}

Next let us verify \eqref{eq-Toepex}. On the fiber $\CM^2$ acts the Cayley transform $c=2^{-\frac{1}{2}}\binom{1 \;-\imath}{1\;\;\;\imath}$. Let us set $\widetilde{O}=(\one\otimes c)\,\widehat{O}\,(\one\otimes c^*)$ and $\widetilde{P}=(\one\otimes c)\,\widehat{P}\,(\one\otimes c^*)$. One can then readily check
$$
\widetilde{O}
\;=\;
\begin{pmatrix}
S & 0 \\ 0 & S^*
\end{pmatrix}
\;,
\qquad
\widetilde{P}
\;=\;
\begin{pmatrix}
p_> & 0 \\ 0 & p_\geq
\end{pmatrix}
\;,
$$
where $p_>=\sum_{n>0}p_n$ and $p_\geq=\sum_{n\geq 0}p_n$. Hence
$$
\widetilde{P}
\widetilde{O}
\widetilde{P}
\;=\;
\begin{pmatrix}
p_>S p_> & 0 \\ 0 & p_\geq S^*p_\geq
\end{pmatrix}
\;.
$$
Both entries are unilateral shifts, one left and one right, so that the kernel is indeed of dimension $1$, and hence that of $POP$ as well.

\section{Examples of $\ZM_2$-spectral flow in topological insulators}
\label{sec-TopIns}

Model Hamiltonians for topological insulators are classified by their symmetry type, see {\it e.g.} \cite{AZ,GS}. In this paper we consider only two such symmetries, particle-hole symmetry (PHS) and time reversal symmetry (TRS). These are described mathematically by conjugate linear involutions or by conjugate linear complex structures.  It is this conjugate linearity that leads to the need for real $K$-theory and hence $\mathbb Z_2$-valued spectral flow.

\vspace{.2cm}

We begin by constructing a Hamiltonian with PHS. Let $T\in\Ff^1(\Hh_\RM)$ be a real skew-adjoint Fredholm operator on $\Hh_\RM$. Its extension as a real linear operator to the complexified Hilbert space $\Hh_\CM=\Hh_\RM\otimes\CM$ is still denoted by $T$. Associated to this $T$, one can define a  Hamiltonian of Bogoliubov-de Gennes (BdG) type \cite{AZ} in the so-called Majorana representation by
\begin{equation}
\label{eq-MajoRep}
H_\Maj\;=\;\imath\,T
\;.
\end{equation}
This is a self-adjoint operator acting on $\Hh_\CM$. The corresponding Atiyah-Singer $\ZM_2$-index \eqref{eq-Z2skew} is
$$
\Ind_1(H_\Maj)
\;=\;
\dim_\CM(\Ker_\CM(H_\Maj))\;\mbox{\rm mod}\;2
\;.
$$
This separates the set of essentially gapped BdG Hamiltonians into two sets, those with an even number of zero (Majorana) modes, and those with an odd number. Now let us furnish $\Hh_\CM$ with some non-trivial grading and let $C$ be the Cayley transformation in that grading
$$
C\;=\;\frac{1}{\sqrt{2}}\begin{pmatrix} 1 & -\imath \\ 1 & \imath \end{pmatrix}
\;.
$$
This then brings $H_\Maj$ into the more conventional BdG form:
$$
H
\;=\;
{C}\,H_\Maj\,C^*
\;.
$$
Recall that for any operator $A$, we set $\overline{A}=\Cc A\Cc$ and $A^t=\overline{A}^*$. The BdG equation is then
\begin{equation}
\label{eq-PHS}
K^*\,\overline{H}\,K\;=\;-\,H\;,
\qquad
K=\begin{pmatrix} 0 & 1 \\ 1 & 0 \end{pmatrix}
\;.
\end{equation}
Now the upper component is interpreted as particles, and the lower one as anti-particles. Given a one-parameter family $t\in[0,1]\mapsto H_t$ of BdG-operators with $0\not\in\sigma_\ess(H_t)$ for all $t\in[0,1]$ and $0\not\in\sigma(H_0)\cup\sigma(H_1)$, one can consider the associated $\ZM_2$-valued spectral flow. In the following, we will construct two examples of non-trivial $\ZM_2$-valued spectral flow in one-dimensional BdG Hamiltonians. One is linked to a flux tube argument allowing us to describe zero modes attached to a defect in the model, the other to a cycle used for orbital polarization.

\subsection{Flux tube through a Kitaev chain}

The infinite and `clean' (meaning no disorder)  Kitaev chain is described by a Hamiltonian on $\ell^2(\ZM)\otimes\CM^2$ given by
\begin{equation}
\label{eq-KitaevDef}
H
\;=\;
\frac{1}{2}\,
\begin{pmatrix}
S+S^*+2\mu & \imath(S-S^*) \\
\imath(S-S^*) & -(S+S^*+2\mu)
\end{pmatrix}
\;.
\end{equation}
Here $S$ denotes the right shift and $\mu\in\RM$ is a chemical potential. Using the Pauli matrices
$$ 
\sigma_1
\;=\;
\begin{pmatrix} 0 & 1 \\ 1 & 0 \end{pmatrix}
\;,
\qquad
\sigma_2
\;=\;
\begin{pmatrix} 0 & -\imath \\ \imath & 0 \end{pmatrix}
\;,
\qquad
\sigma_3
\;=\;
\begin{pmatrix} 1 & 0 \\ 0 & -1 \end{pmatrix}
\;,
$$
one has
\begin{equation}
\label{eq-S0def}
H\;=\;
S_0+S_0^*+\mu\,\one\otimes\sigma_3
\;,
\qquad
S_0\,=\,S\otimes\frac{1}{2}(\sigma_3+\imath\,\sigma_1)\,=\,
S\otimes\frac{1}{2}\begin{pmatrix} 1 & \imath \\ \imath & -1 \end{pmatrix}
\;.
\end{equation}
The operator $S_0$ is the right shift on the line with particle-hole fiber, with coupling terms going from the particle to the hole fiber and visa versa. The index $0$ indicates that there is no flux  pushed through. Below, the definition will be extended to $S_\alpha$ with a flux $2\pi\alpha$. For $\mu\not=\{-1,1\}$ one has $0\not\in\sigma(H)$. 

\vspace{.2cm}

The Hamiltonian $H$ has an even PHS (that is, the symmetry is given by an involution and thus its square is the identity) with $K=\sigma_1$ as in \eqref{eq-PHS} and an even TRS (again the operator giving the TRS symmetry squares to the identity) with $J=\sigma_3$. Consequently, in the usual symmetry classification scheme \cite{AZ}, this model lies in the Class BDI. As it also has an induced `chiral symmetry' given by the composition of the PHS and TRS, it has a well-defined $\ZM$-index \cite{PS2}. This reduces to a $\ZM_2$ index  if the TRS is broken and the system then only lies in the Class D \cite{GS}. This $\ZM_2$ index is equal to the number of zero modes (Majorana states) that a half-space restriction of the model has, taken modulo $2$.  All these facts are stable under perturbations by disorder, as long as the stated symmetries hold. For sake of simplicity, we will first work with the clean model and then add perturbations by homotopy at the end of the section.

\vspace{.2cm}

\begin{figure}
\centering
  \begin{minipage}[b]{0.4\textwidth}
\includegraphics[height=1.65cm]{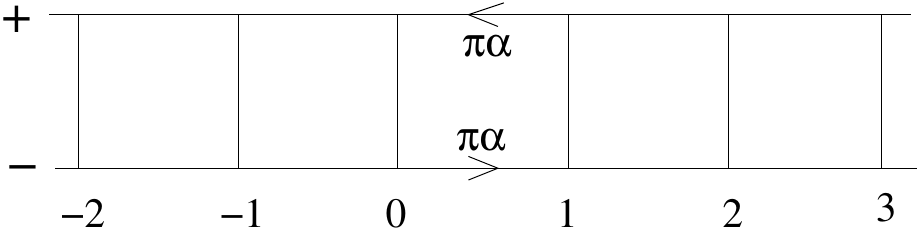}
  \end{minipage}
\hspace{0.5cm}
 \begin{minipage}[b]{0.4\textwidth}
\includegraphics[height=1.6cm]{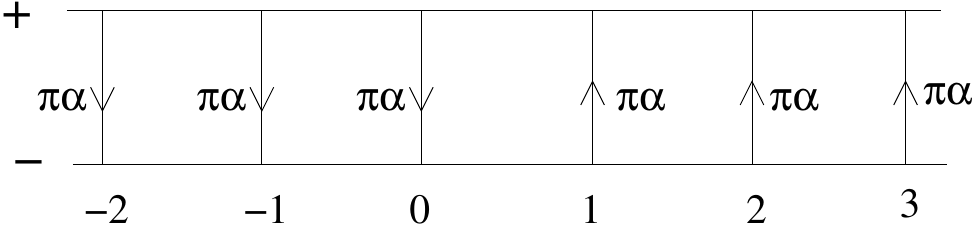}
  \end{minipage}
\caption{\it 
The two gauges $A$ and $\widetilde{A}$ used for a flux $2\pi\alpha$ through the cell between $0$ and $1$.
\label{fig-gauges}
}
\end{figure}

Now let us consider the model as defined on a (discrete) strip $\ZM\times\{+,-\}$ of width $2$. A magnetic flux $2\pi\alpha\in[0,2\pi)$ will be inserted in the cell between $0\in\ZM$ and $1\in\ZM$. This can be realised by various gauge potentials, namely a real-valued function on the oriented links of the underlying lattice $\ZM\times\{+,-\}$. It is given by a function $A:(\ZM\times\{+,-\})^2\to\RM$  which is anti-symmetric in its two arguments and is only non-vanishing for two neighbouring  sites of the lattice $\ZM\times\{+,-\}$  (see Section~2.1 of \cite{DS} for a concise review of the basic facts on gauges and gauge transformations used below). We will actually use two different gauges which are  represented graphically in Fig.~\ref{fig-gauges}. The first gauge is
$$
A((n,\eta),(n',\eta'))
\;=\;
\pi\alpha\, \delta_{n,1}\delta_{n',0}\delta_{\eta,-}\delta_{\eta',-}
\;-\;
\pi\alpha\, \delta_{n,1}\delta_{n',0}\delta_{\eta,+}\delta_{\eta',+}
\;,
$$
where $(n,\eta),(n',\eta')\in \ZM\times\{+,-\}$. Following again \cite{DS}, the Hamiltonian with a flux tube is in this gauge given by
\begin{equation}
\label{eq-Salpha}
H_\alpha
\;=\;
S_\alpha+S_\alpha^*+\mu\,\one\otimes\sigma_3
\;,
\qquad
S_\alpha\;=\;S_0\,+\,\pi_1(\pi_0)^*\otimes
\frac{1}{2}
\begin{pmatrix}
e^{-\imath\pi\alpha}-1 & \imath(e^{\imath\pi\alpha}-1)
\\
\imath(e^{-\imath\pi\alpha}-1) & -(e^{\imath\pi\alpha}-1)
\end{pmatrix}
\;,
\end{equation}
where the partial isometry $\pi_n$ onto site $n$ are defined as in Section~\ref{sec-ex}.
The second (non-local) gauge is
$$
\widetilde{A}((n,\eta),(n',\eta'))
\;=\;
\pi\alpha\, \delta_{n,n'}\delta_{n\leq 0}\delta_{\eta,-}\delta_{\eta',+}
\;-\;
\pi\alpha\, \delta_{n,n'}\delta_{n\geq 1}\delta_{\eta,-}\delta_{\eta',+}
\;,
$$
and the Hamiltonian
\begin{equation}
\label{eq-Salphatilde}
\widetilde{H}_\alpha
\;=\;
\widetilde{S}_\alpha+\widetilde{S}_\alpha^*+\mu\,\one\otimes\sigma_3
\;,
\qquad
\;\widetilde{S}_\alpha\;=\;\frac{1}{2}
\begin{pmatrix}
\one & \imath\,e^{\imath\pi\alpha\,\sgn'(X)}
\\
\imath\,e^{-\imath\pi\alpha\,\sgn'(X)} & -\one
\end{pmatrix}
\cdot
S\otimes\one_2
\;,
\end{equation}
where $\sgn'(X)=\sgn(X)\,+\,p_0$ is the modified sign of the position operator with $\sgn'(X)\pi_0=\pi_0$ ({\it cf.} Section~\ref{sec-ex}).
First of all, let us note that both $\alpha\mapsto H_\alpha$ and $\alpha\mapsto\widetilde{H}_\alpha$ are paths in Hamiltonians with the even BdG symmetry \eqref{eq-PHS}, and thus after Cayley transform paths in $\Ff^1(\Hh_\RM)$. Second of all, as $A$ and $\widetilde{A}$ induce the same magnetic field, there exists a so-called gauge transformation $G:\ZM\times\{-,+\}\to{\RM}$  such that, for $|(n,\eta)-(n',\eta')|=1
$, 
\begin{equation}
\label{eq-gaugetrafo}
\widetilde{A}((n,\eta),(n',\eta'))\;=\;A((n,\eta),(n',\eta'))+G(n,\eta)-G(n',\eta')
\;, 
\end{equation}
and via this gauge transformation the Hamiltonians are unitarily equivalent:
\begin{equation}
\label{eq-gaugetraf}
\widetilde{H}_\alpha
\;=\;
e^{\imath G(X,\eta)}\,H_\alpha\,e^{-\imath G(X,\eta)}
\;.
\end{equation}
Let us collect a few facts related to these Hamiltonians.

\begin{prop}
\label{prop-listprop}
The following holds for all $\alpha\in\RM$ with $K=\one\otimes\sigma_1$ and $I=\one\otimes\sigma_3$.
\begin{enumerate}[{\rm (i)}]

\item $K^*\Cc{H}_\alpha\Cc K=-H_\alpha$ and $\sigma(H_\alpha)=-\sigma(H_\alpha)$

\item $K^*\Cc\widetilde{H}_\alpha\Cc K=-\widetilde{H}_\alpha$ and $\sigma(\widetilde{H}_\alpha)=-\sigma(\widetilde{H}_\alpha)$

\item $H_\alpha-H_0$ is compact and $\sigma_\ess(H_\alpha)=\sigma_\ess(H_0)$

\item $\sigma(H_\alpha)=\sigma(\widetilde{H}_\alpha)$

\item $H_{\alpha+2}=H_\alpha$

\item $\widetilde{H}_1=I^* \widetilde{H}_0 I$ and  $\sigma(\widetilde{H}_1)=\sigma( \widetilde{H}_0 )$

\item $\Cc \widetilde{H}_\alpha\Cc= \widetilde{H}_{1-\alpha}$ and $\sigma(H_\alpha)=\sigma(H_{1-\alpha})$

\item ${H}_1+{H}_0 =\widehat{H}_l\oplus\widehat{H}_r$ where $\widehat{H}_l$ and $\widehat{H}_r$ are the restrictions of ${H}_0$ to the left and right half-line Hilbert spaces $\ell^2(\NM_{\leq 0})\otimes\CM^2$ and $\ell^2(\NM_{\geq 1})\otimes\CM^2$  respectively, both with Dirichlet boundary conditions.

\item For $|\mu|<1$, $\dim_\CM(\Ker_\CM(\widehat{H}_l))=\dim_\CM(\Ker_\CM(\widehat{H}_r))=1$ and $\dim_\CM(\Ker_\CM({H}_0+{H}_1))=2$

\end{enumerate}

\end{prop}

\noindent {\bf Proof.} Items (i) and (ii) can be checked algebraically from \eqref{eq-Salpha} and \eqref{eq-Salphatilde}, and (iii) follows from \eqref{eq-Salpha} combined with Weyl's theorem of the essential spectrum. Furthermore, (iv) results from the gauge transformation \eqref{eq-gaugetraf} and (v) is obvious. Items (vi) and (vii) are due to the identity
$$
e^{\imath \pi\,\sgn'(X)}\;=\;e^{-\imath \pi\,\sgn'(X)}\;=\;-\;\one\;.
$$
For (viii), one uses
$$
S_1\;=\;S_0\;+\;\pi_1(\pi_0)^*|\otimes
\begin{pmatrix}
-1 & -\imath \\ -\imath & 1 \end{pmatrix}
\;.
$$
Comparing with \eqref{eq-S0def}, one sees that the second summand reverses the links between $0\in\ZM$ and $1\in\ZM$, so that $S_0+S_1$ is indeed a direct sum of two uni-lateral shifts (tensorized with a $2\times 2$ matrix). Finally (ix) results from the fact that both $\widehat{H}_l$ and $\widehat{H}_r$ have exactly one Majorana boundary state for $|\mu|<1$, see Section~3.6.1 of \cite{GS}.
\hfill $\Box$

\vspace{.2cm}

Now we can consider the $\ZM_2$-valued spectral flow along the path $\alpha\in[0,1]\mapsto H_\alpha$, strictly speaking to the path of real skew-adjoints $\alpha\in[0,1]\mapsto T_\alpha=-\imath\, C^*H_\alpha C$ of the associated Majorana representation \eqref{eq-MajoRep}.

\begin{prop}
\label{prop-fluxflow} 
For $|\mu|<1$,
$$
\SF_2(\alpha\in[0,1]\mapsto H_\alpha)
\;=\;
1\;.
$$
\end{prop}

\noindent {\bf Proof.} Let us first argue that
$$
\SF_2(\alpha\in[0,1]\mapsto H_\alpha)
\;=\;
\SF_2(t\in[0,1]\mapsto H(t)=(1-t)H_0+tH_1)
\;.
$$
As $\alpha\in[0,1]\mapsto H_\alpha$ and the straight line path $H(t)=(1-t)H_0+tH_1$ have the same end points, this follows from the homotopy invariance (Theorem~\ref{theo-hmot_inv}) once we have provided a homotopy between the two paths. But $H_\alpha=H_0+L_\alpha$ for some finite range $L_\alpha$, see \eqref{eq-Salpha}, and on the other hand $H(t)=H_0+L'_t$ for another finite range $L'_t$. Hence a homotopy of paths with fixed end points is given by $s\in[0,1]\mapsto H(t,s)=H_0+sL_t+(1-s)L'_t$. In conclusion, it remains to show $\SF_2(t\in[0,1]\mapsto H(t))=1$. For all $|\mu|<1$, the zero energy $0$ lies in a gap of $H_0$ and $H_1$. Hence there is no $\ZM_2$-spectral flow when $\mu$ is homotopically deformed to $\mu=0$, so that it is sufficient to consider this case. But for $\mu=0$, one can check starting from \eqref{eq-KitaevDef} that $(H_0)^2=\one$. By Proposition~\ref{prop-listprop} and the gauge transformation \eqref{eq-gaugetraf} one also concludes $(H_1)^2=\one$. Therefore the real skew-adjoints associated to the Majorana representation \eqref{eq-MajoRep} are actually complex structures, so that one can invoke Proposition~\ref{prop-basicSF} to calculate $\SF_2(t\in[0,1]\mapsto H(t))=\SF_2(H_0,H_1)$ using  Proposition~\ref{prop-listprop}(ix). 
\hfill $\Box$

\vspace{.2cm}

The main application will concern the spectral properties at $\alpha=\frac{1}{2}$. This is particularly interesting because the Hamiltonian at $\alpha=\frac{1}{2}$ has the even TRS $\Cc\widetilde{H}_{\frac{1}{2}}\Cc=\widetilde{H}_{\frac{1}{2}}$. It thus lies in the same universality class BDI as the Kitaev model, and is, in fact, obtained from the latter model by the perturbation of a local defect at the sites $0\in\ZM$ and $1\in\ZM$. The following result shows that this defect necessarily has zero modes attached to it. As $\Ind_1(H_\alpha)=0$ for all $\alpha$, these zero modes are always evenly degenerate.

\begin{theorem}
\label{theo-defect}
For $|\mu|<1$, the Hamiltonian ${H}_{\frac{1}{2}}$ has an odd number of evenly degenerate zero eigenvalues:
$$
\tfrac{1}{2}\,\dim_\CM(\Ker_\CM({H}_{\frac{1}{2}}))\;\mbox{\rm mod}\,2
\;=\;1
\;.
$$
\end{theorem}

\noindent {\bf Proof.} The symmetry property $\Cc \widetilde{H}_\alpha\Cc= \widetilde{H}_{1-\alpha}$  Proposition~\ref{prop-listprop}(vii) combined with the last claim of Theorem~\ref{theo-welldef} shows that for any $\alpha_0<\frac{1}{2}$ such that $H_{\alpha_0}$ has minimal kernel dimension,
$$
\SF_2(\alpha\in[0,\alpha_0]\mapsto H_\alpha)
\;=\;
\SF_2(\alpha\in[1-\alpha_0,1]\mapsto H_\alpha)
\;.
$$
Hence from the concatenation property
$$
\SF_2(\alpha\in[0,1]\mapsto H_\alpha)
\;=\;
\SF_2(\alpha\in[\alpha_0,1-\alpha_0]\mapsto H_\alpha)
\;.
$$
As $H_\alpha$ is analytic in $\alpha$, one can conclude that at $\alpha=\frac{1}{2}$ there is an odd number of eigenvalue crossings by Proposition~\ref{prop-fluxflow}. 
\hfill $\Box$

\vspace{.2cm}

Finally, let us comment on adding disorder or other perturbations to the Hamiltonian:
$$
H_\alpha(\lambda)
\;=\;
H_\alpha\,+\,\lambda\,V
\;.
$$ 
Here $V$ is a possibly random perturbation representing disorder in the system but having the BdG and TRS symmetries $K^*\overline{V}K=-V$ and $I^*\overline{V}I=V$. Furthermore, $\lambda$ is chosen to be sufficiently small such that the gap in the essential spectrum of $H_0$ does not close. This essentially imposes that $|\lambda|\leq C\,|\mu|$ for some suitable constant depending on the norm of $V$. Now all statements proved above directly transpose to $H_\alpha(\lambda)$, in particular, also Theorem~\ref{theo-defect}.

\subsection{$\ZM_2$-polarization}

This short last section indicates that there are further natural instances where the $\ZM_2$-valued spectral flow appears. It is based on several prior results which are only briefly described and the reader is urged to read up in the cited references.  Let us begin by recalling from \cite{PS2} the connection between polarization and spectral flow of half-sided operators based on the bulk-boundary correspondence. We restrict to dimension $1$, even though this connection holds in any dimension. Suppose given a smooth loop $t\in[0,2\pi)\mapsto h_t$ of periodic Hamiltonians on $\ell^2(\ZM)\otimes \CM^N$. The orbital polarization at Fermi level $\mu$ lying in a gap of $h_t$ for all $t$ is defined by
$$
\Delta P
\;=\;
\imath\;\int^{2\pi}_0dt\,\TR_N\left(\pi_0^* p_t\big[\partial_tp_t,\imath [X,p_t]\big]\pi_0\right)
\;,
$$
where $p_t=\chi(h_t\leq \mu)$ is the (instanteneous) Fermi projection. Actually, the polarization $\Delta P$ is defined as the accumulated charge during one loop and then the above is the so-called King-Smith-Vanderbilt formula which holds in the adiabatic limit as proved in \cite{ST}. 

\vspace{.2cm}

The second important fact for the following is that $\Delta P$ is $2\pi$ times an integer called the Chern number $\mbox{\rm Ch}(p)$ where $p=(p_t)_{t\in[0,2\pi)}$. Again this Chern number is also given as the index of a certain Fredholm operator \cite{PS2}. Next let us consider the restrictions $\widehat{h}(t)$ of $h(t)$ to $\ell^2(\NM)\otimes\CM^N$, say, with Dirichlet boundary conditions (actually any local boundary condition will do). These operators may have bound states and it is proved in \cite{PS2} that the standard complex spectral flow of these eigenvalues through $\mu$ is equal to the Chern number, namely
$$
\Delta P\;=\;2\pi\,\SF\big(t\in[0,2\pi)\mapsto \widehat{h}_t\;\mbox{\rm by } \mu \big)
\;.
$$
Based on these facts, we now construct an example of a BdG-Hamiltonian with non-trivial $\ZM_2$-valued spectral flow. On $\ell^2(\ZM)\otimes\CM^{2N}$, let us set
$$
H_t(\lambda)
\;=\;
\begin{pmatrix}
h_t & 0 \\ 0 &-\overline{h}_t 
\end{pmatrix}
\;+\;
\lambda\,V_t
\;,
$$
where $\lambda\geq 0$ is a coupling constant and $t\in[0,2\pi)\mapsto V_t$ is a smooth loop of bounded BdG operators, namely $K^*\overline{V_t}K=-V_t$. Then also $H_t(\lambda)$ satisfies the BdG equation \eqref{eq-PHS}. In the following, it will be assumed that $\lambda$ is sufficiently small such that $0\not\in\sigma(H_t(\lambda))$ for all $t\in[0,2\pi)$. Due to the minus sign, the polarization $\Delta P(\lambda)$ of $t\in[0,2\pi)\mapsto H_t(\lambda)$ vanishes and is equal to the spectral flow of the half-line restriction $\widehat{H}_t(\lambda)$ to $\ell^2(\NM)\otimes\CM^{2N}$:
$$
\Delta P(\lambda)
\;=\;
2\pi\,\SF\big(t\in[0,2\pi)\mapsto \widehat{H}_t(\lambda)\big)
\;=\;0\;.
$$
Note that the vanishing of the spectral flow on the r.h.s. also follows immediately from the BdG symmetry which implies that the spectrum of $\widehat{H}_t(\lambda)$ is always symmetric around $0$ so that the spectral flow necessarily vanishes. For  non-trivial $V_t$ and say intermediate $\lambda\not=0$ (still such that the gap remains open) it may not be possible to define the spectral flow of the upper left and lower left components separately any more. Nevertheless the $\ZM_2$-valued spectral flow may be topologically non-trivial, as shows the following result.

\begin{theorem}
\label{theo-BdGflow}
For $\lambda$ sufficiently small, the $\ZM_2$-valued spectral flow satisfies
$$
\SF_2\big(t\in[0,2\pi)\mapsto \widehat{H}_t(\lambda)\big)
\;=\;
\tfrac{1}{2\pi}\,\Delta P\;\,\mbox{\rm mod }2\;\in\;\ZM_2
\;.
$$
This quantity is called the $\ZM_2$-polarization. In particular, if it is equal to $1$, there is at least one $t\in (0,2\pi)$ such that $\widehat{H}_t$ has a zero mode with multiplicity equal to $2$ modulo $4$. 
\end{theorem}


\end{document}